# Introduction to Mathematical Programming-Based Error-Correction Decoding

**A tutorial**

Michael Helmling

August 31, 2015

# Contents

# Preface

This tutorial-style introduction into the topic of *error-correction decoding based on mathematical programming*, including the most prominent application called *LP decoding*, has emerged from the introductory chapter of my PhD thesis [Hel15]. However, I have a priori intended to publish this introduction after my graduation, hoping that it might be useful to students and researchers beyond those who happen to come across my thesis – especially because there is to date, as far as I am aware, no introductory textbook on "LP decoding" that covers recent research, which makes it hard to get into the topic for newcomers. The best starting point so far is probably still Feldman's thesis, which has become a little long in the tooth now.

Of course, the text at hand is as well biased to the specific topics that I have worked with. My hope is, however, that others (i.e., you!) will help to improve and extend the document by contributing bugfixes, new sections, examples, proofs or whatever you fell is missing. To that end, the LaTeX sources are publicly available via GitHub[1], and I encourage everyone to contribute via bug reports or pull requests.

Michael Helmling<br>Koblenz, Germany<br>2015

[1] https://github.com/supermihi/lpdintro

# Introduction

Decoding error-correctiong codes by methods of mathematical optimization, most importantly linear programming, has become an important alternative approach to both algebraic and iterative decoding methods since its introduction by Feldman, Wainwright, and Karger [FWK05]. At first celebrated mainly for its analytical powers, real-world applications of LP decoding are now within reach thanks to most recent research ([LDR12; Gen+14]). This document gives an elaborate introduction into both *mathematical optimization* (Chapter 2) and *coding theory* (Chapter 3) as well as a review of the contributions by which these two areas have found common ground in Chapter 4.

We have nevertheless made it our aim that a reader who is familiar with mathematics in general but has no specific proficiency in either optimization or coding theory will be able to comprehend the most important ideas and concepts of the matter. Being originated as a thesis introduction, the presentation might still be rather compact as compared to a textbook. However, you can change that by either contributing more elaborate explanations or examples directly, or by asking someone else (e.g. me) to do so.

The assumptions we make as to the reader's mathematical background mainly consist of undergraduate linear algebra. To a smaller extent, familiarity with graph theory and algorithmic concepts will be helpful, and occasionally we will encounter probabilities. Notation and nomenclature for these background topics are fixed in Chapter 1.

# 1 Notation and Preliminaries

## Notation

We use the symbols $\mathbb{N}$ and $\mathbb{Z}$ for the natural (starting with 1) and integral numbers, respectively, $\mathbb{F}_2$ for the binary field (which is sometimes called GF(2)), $\mathbb{Q}$ for the rational numbers and $\mathbb{R}$ for the reals. If $x \in \mathbb{R}$, then $\lfloor x \rfloor$ and $\lceil x \rceil$ denote the largest integer $\leq x$ and the smallest integer $\geq x$, respectively.

For a set $R$ and $n \in \mathbb{N}$, $R^n$ denotes the $n$-dimensional vector space over $R$ if $R$ is a field, and the set of $n$-tuples of elements of $R$ otherwise. In either case, the $n$-tuples are called *vectors* and operations like addition or comparison, whenever applicable, are understood element-wise: if $x = (x_1, \dots, x_n)$ and $y = (y_1, \dots, y_n)$, then

$$x + y = (x_1 + y_1, \dots, x_n + y_n)$$

and

$$x \leq y \text{ if and only if } x_i \leq y_i \text{ for } i = 1, \dots, n.$$

## Matrices and Vectors

By $X^{m \times n}$ we denote the set of matrices with $m$ rows and $n$ columns and entries from the set $X$. For a matrix $A$, by $A_{i,j}$ we denote the element at the $i$-th row and $j$-th column of $A$, $A_{i,\bullet}$ and $A_{\bullet,j}$ are *the* $i$-th row and $j$-th column, respectively, and the transpose of $A$ is the $n \times m$ matrix $A^T$ with entries $A^T_{j,i} = A_{i,j}$. Throughout the text we regard vectors $x \in X^n$ as *column* vectors, i.e., identify them with $k \times 1$ matrices. We sometimes use *(ordered) index sets* instead of individual indexes: for instance, if $x \in X^n$ and $I = (i_1, \dots, i_k)$, where $\{i_1, \dots, i_k\} \subseteq \{1, \dots, n\}$, then $x_I = (x_{i_1}, \dots, x_{i_k})$. The same notation is used for row and column indexing, respectively, of matrices.

When $A \in R^{m \times n}$ is an $m \times n$ matrix over a ring $R$, an *elementary row operation* on $A$ is one of the following operations: *(a)* multiply a row of $A$ by a scalar: $A_{i,\bullet} \leftarrow sA_{i,\bullet}$, or *(b)* replace a row by the weighted sum of itself and another row: $A_{i,\bullet} \leftarrow sA_{i,\bullet} + tA_{j,\bullet}$. If $R$ is a field, any finite sequence of elementary row operations (taking the scalars from $R$) leaves the range $\{Ax \colon x \in R^n\}$ of $A$ unaltered. Such a sequence is called a *Gaussian pivot* on $A_{i,j}$ if it turns the $j$-th column of $A$ into the $i$-th unit vector (see Figure 1.1), and *Gaussian elimination* if it changes a submatrix of $A$ into a triangular or diagonal matrix (the terms Gauss-Jordan pivot and, for diagonalization, Gauss-Jordan elimination are also used).

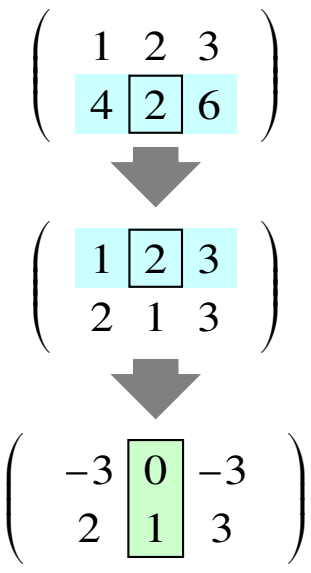

Figure 1.1: Example Gaussian pivot operation.

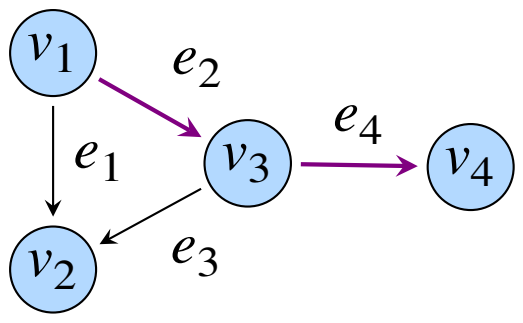

Figure 1.2: An acyclic directed graph with the path $P = (v_1, v_3, v_4)$ (using vertices) or $P = (e_2, e_4)$ (using edges) highlighted.

## Graphs

A *graph* $G = (V, E)$ is defined by a finite set $V$, called the *nodes* or *vertices* of $G$, and a set $E \subseteq V \times V$ of *edges* or *arcs* of $G$. There are both *undirected* graphs, in which an edge $(u, v) \in E$ is identified with the unordered set $\{u, v\}$, and *directed graphs*, where $(u, v)$ is perceived as an ordered pair. For a node $v \in V$ of a directed graph, we define $\delta^+(v) = \{(v, u) : (v, u) \in E\}$ and $\delta^-(v) = \{(u, v) : (u, v) \in E\}$ as the *outbound* and *inbound* edges, respectively, of $v$.

A finite sequence $P = (v_1, \dots, v_k) \in V^k$ is called a $v_1$–$v_k$ *path*, or simply *path*, of $G$, if $(v_i, v_{i+1}) \in E$ for $i = 1, \dots, k-1$. Each path can alternatively be defined by a sequence of edges, $P = (e_1, \dots, e_{k-1})$, where $e_i = (v_i, v_{i+1})$ for $i = 1, \dots, k-1$. The path $P$ is called a *cycle* if $v_1 = v_k$ (see Figure 1.2 for an example). A graph is *acyclic* if no cycle exists.

For a graph $G = (V, E)$ and $V' \subseteq V$, the *subgraph induced by* $V'$ is the graph $G' = (V', E')$ where $E' = \{(u, v) \in E : u \in V' \text{ and } v \in V'\}$ consists of all edges that connect two vertices of $V'$. Finally, a graph is called *bipartite* if there is a partition $V = V_1 \dot\cup V_2$ such that $E \subseteq \{V_1 \times V_2\} \cup \{V_2 \times V_1\}$, i.e., no edge connects two vertices of the same set $V_i$, $i = 1, 2$.

## Complexity

The symbols P and NP are used to denote the complexity classes of problems that are *solvable* (P) and *verifiable* (NP) in polynomial time, and a problem is called *NP-hard* if it is "at least as hard" as *every* problem in NP, i.e., each problem in NP can be reduced to it in polynomial time. The *Landau notation* $f(n) = O(g(n))$ states that the asymptotic growth rate of $f(n) : \mathbb{N} \to \mathbb{R}$ is upper bounded by $g(n)$, i.e., there exist $M > 0$ and $n_0 \in \mathbb{N}$ such that $f(n) \le Mg(n)$ for $n > n_0$. Note that a thorough understanding of complexity theory is not a prerequisite to reading this text.

## Probability

Since we only come across basic probability calculations and they are not central to our text, we use simplified notation. Namely, we frequently denote both a random variable and its outcomes by the same symbol, say $x$, and use $P(x)$ for the probability mass or density function of $x$, whatever applies—the exact meaning of the symbols will always become clear from the context. If $y$ is another random variable, $P(x \mid y_0)$ is the conditional probability function of $x$ given the event $\{y \in y_0\}$. Similarly, $P(x_0 \mid y)$ is called the likelihood function of $y$, given $\{x \in x_0\}$. *Bayes' theorem* states that $P(x_0 \mid y_0) = \frac{P(y_0|x_0)P(x_0)}{P(y_0)}$ for any two events $x_0$ of $x$ and $y_0$ of $y$, provided $P(y_0) \neq 0$.

# 2 Optimization Background

Mathematical optimization is a discipline of mathematics that is concerned with the solution of problems arising from mathematical models that typically describe real-world problems, e.g. in the areas of transportation, production planning, or organization processes. These models are generally of the form

$$\begin{aligned} \min \quad & f(x) \\ \text{subject to (s.t.)} \quad & x \in X, \end{aligned}$$

where $X \subseteq \mathbb{R}^n$, for some $n \in \mathbb{N}$, is the *feasible set* and $f: X \to \mathbb{R}$ the *objective function* that evaluates a feasible point $x \in X$; $f(x)$ is called the *objective value* of $x$. An $x^* \in X$ minimizing the objective function is called an *optimal solution*, the corresponding value $z^* = f(x^*)$ the *optimal objective value*. If $X = \emptyset$, the problem is said to be *infeasible* and we define $z^* = \infty$ in that case. If on the other hand $f(x)$ is unbounded from below among $X$, we define $z^* = -\infty$.

The theory of optimization further subdivides into several areas that depend on the structure of both $f$ and $X$. Within this text we will encounter three major problem classes: linear programs (LPs), integer linear programs (IPs), and combinatorial optimization problems, which are often modeled by means of an IP.

A common ground in the analysis of these three types of problems is the *polyhedral structure* of the feasible set. The part of polyhedral theory that is necessary for this text is reviewed in Section 2.1. For an LP, the feasible set is a polyhedron that is given explicitly by means of a defining set of linear inequalities, which makes these problems relatively easy to solve. LPs and the *simplex method*, the most important algorithm to solve them, are covered in Section 2.2.

In contrast to LPs, the feasible region of an IP is given only implicitly as the set of *integral* points within a polyhedron. Although the result exhibits again a polyhedral structure (as long as finding an optimal solution is the concern), IPs are much harder to solve than LPs; in fact, integer programming in general is an NP-hard optimization problem. Some theoretical foundations and techniques to nonetheless tackle such problems are collected in Section 2.3.

Proofs, examples and detailed explanations are widely omitted in this chapter. For a very exhaustive and detailed textbook on linear programming and the simplex method, see [Dan63]. A complete and rigorous yet challenging reference for linear and integer programming is the book by Schrijver [Sch86]. A well-written algorithm-centric introduction to linear, integer and also nonlinear optimization can be found in [FKS10]. Finally, for integer and combinatorial optimization we refer to [NW88].

## 2.1 Polyhedral Theory

The mathematical objects that we talk about in this section live in the $n$-dimensional Euclidean space $\mathbb{R}^n$. Before we begin to discuss polyhedra and the theory around them, we briefly rush through some basic concepts which are necessary for that task.

### 2.1.1 Convex Sets and Cones

*Convexity* is one of the most important concepts in mathematical optimization: intuitively speaking, a geometric object is convex if the straight line between any two points of the object lies completely inside of it.

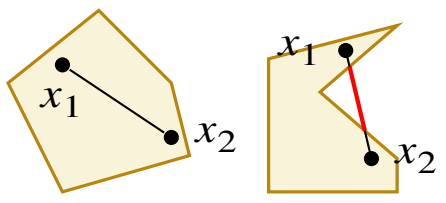

Figure 2.1: A convex and a nonconvex set.

**2.1 Definition (convex and conic sets and hulls):** A set $X \subseteq \mathbb{R}^n$ is called *convex* if for any $x_1, x_2 \in X$ and $\lambda \in [0, 1]$ also $\lambda x_1 + (1 - \lambda)x_2 \in X$.

A *convex combination* of $X$ is a sum of the form

$$\sum_{x \in X} \lambda_x x\colon \ \lambda_x \geq 0 \text{ for all } x \in X \text{ and } \sum_{x \in X} \lambda_x = 1, \tag{2.1}$$

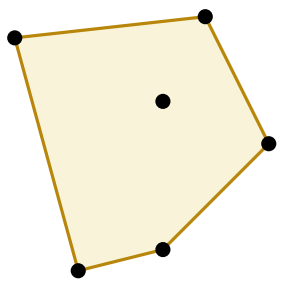
Figure 2.2: Convex hull of six points.

where in the case of an infinite $X$ we assume that almost all $\lambda_x = 0$ so that the above expression makes sense. The *convex hull* of $X$ is the smallest convex set containing $X$ or, alternatively, the set of all convex combinations of elements of $X$. ◁

An important class of convex sets is the one of convex *cones*.

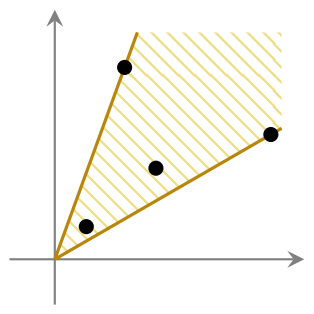
Figure 2.3: Conic hull of four points.

**2.2 Definition (convex cones):** $X \subseteq \mathbb{R}^n$ is called a (convex) *cone* if for any $x_1, x_2 \in X$ and $\lambda_1, \lambda_2 \geq 0$ also $\lambda_1 x_1 + \lambda_2 x_2 \in X$. A *conic combination* is defined like (2.1) but without the condition $\sum_{x \in X} \lambda_i = 1$. Analogously to the above, the *conic hull* $\operatorname{conic}(X)$ is the smallest cone containing $X$, or the set of all conic combinations of elements of $X$. ◁

Geometrically, the conic hull is the largest set that "looks the same" as $\operatorname{conv}(X)$, from the perspective of an observer that sits at the origin.

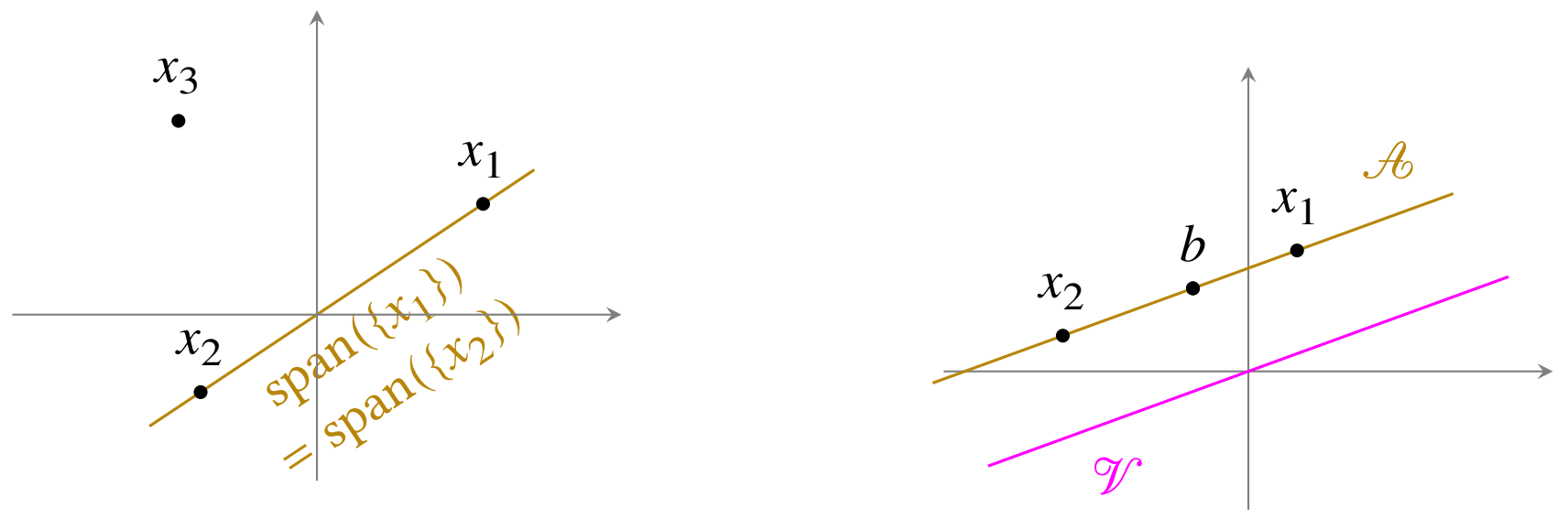

(a) $x_1$ and $x_2$ are linearly dependent, but both are pairwise linearly independent with $x_3$.

(b) Affine space $\mathscr{A} = b + \mathscr{V}$ of dimension 1; $x_1, x_2 \in \mathscr{A}$ are affinely independent.

Figure 2.4: Examples of linear and affine (in)dependence.

### 2.1.2 Affine Spaces

Recall the notion of *linear independence*: a set of points $X = \{x_1, \dots, x_k\} \subseteq \mathbb{R}^n$ is *linearly independent* if no element $x$ of $X$ is contained in the linear subspace that is spanned by the remainder $X \setminus \{x\}$ or, equivalently, if $\operatorname{span}(X') \neq \operatorname{span}(X)$ for all $X' \subsetneq X$ (Figure 2.4(a)). Here, the *span* or *linear hull* of a set $X \subseteq \mathbb{R}^n$ means the smallest linear subspace containing $X$:

$$\operatorname{span}(X) = \bigcap \{\mathscr{V} : \mathscr{V} \text{ is a subspace of } \mathbb{R}^n \text{ and } X \subseteq \mathscr{V}\}\,.$$

The linear hull has an alternative algebraic characterization by means of linear combinations,

$$\operatorname{span}(X) = \left\{ \sum_{i=1}^{k} \lambda_i x_i \colon \lambda_i \in \mathbb{R} \text{ for all } i \in \{1, \dots, k\} \right\}, \tag{2.2}$$

which gives rise to the well-known algebraic formulation of linear independence: $X$ is linearly independent if and only if the system

$$\sum_{i=1}^{k} \lambda_i x_i = 0 \tag{2.3}$$

has the unique solution $\lambda_1 = \dots = \lambda_k = 0$. To see this, note that if there was some $j \in \{1, \dots, k\}$ such that $0 \neq x_j \in \operatorname{span}(X \setminus \{x_j\})$, then (2.2) gave rise to a non-zero solution of (2.3).

For the study of polyhedra, we need the concepts of *affine* hulls and independence, respectively, which is centered around the notion of an *affine subspace* in a very similar fashion as for the linear case above.

**2.3 Definition (affine spaces, hulls, and independence):** A set $\mathscr{A} \subseteq \mathbb{R}^n$ is an *affine subspace* of $\mathbb{R}^n$ if there exists a linear subspace $\mathscr{V} \subseteq \mathbb{R}^n$ and some $b \in \mathbb{R}^n$ such that

$$\mathscr{A} = b + \mathscr{V} = \{b + v \colon v \in \mathscr{V}\}\,. \tag{2.4}$$

The *affine hull* $\operatorname{aff}(X)$ of a set $X = \{x_1, \dots, x_k\} \subseteq \mathbb{R}^n$ is the smallest affine subspace containing $X$. The set $X$ is called *affinely independent* if no $x \in X$ fulfills $x \in \operatorname{aff}(X \setminus \{x\})$ or, equivalently, if $\operatorname{aff}(X) \neq \operatorname{aff}(X')$ for all $X' \subsetneq X$.

Finally, the *dimension* $\dim(\mathscr{A})$ of $\mathscr{A}$ in (2.4) is defined to be the dimension of $\mathscr{V}$. ◁

An affine subspace can thus be envisioned as a linear space that has been translated by a vector (see Figure 2.4(b) for an example). This principle is reflected by the algebraic characterization of the affine hull of $X = \{x_1, \dots, x_k\} \subseteq \mathbb{R}^n$ by means of *affine combinations*: first, move to an arbitrary vector of $X$ (without loss of generality, let this be $x_1$), then add any linear combination of the *directions* from $x_1$ to the other $k-1$ elements of $X$:

$$\operatorname{aff}(X) = \left\{ x_1 + \sum_{i=2}^{k} \lambda_i (x_i - x_1) \colon \lambda_i \in \mathbb{R} \text{ for all } i \in \{2, \dots, k\} \right\};$$

by defining $\lambda_1 = 1 - \sum_{i=2}^{k} \lambda_i$ one can easily derive the equivalent definition

$$\operatorname{aff}(X) = \left\{ \sum_{i=1}^{k} \lambda_i x_i \colon \lambda_i \in \mathbb{R} \text{ for all } i \in \{1, \dots, k\} \text{ and } \sum_{i=1}^{k} \lambda_i = 1 \right\}.$$

An algebraic definition of affine dependence can be derived in exactly the same way as for linear dependence: $x_j \in \operatorname{aff}\left(X \setminus \{x_j\}\right)$ if and only if one can write $x_j$ as

$$x_j = x_l + \sum_{i \neq j, l} \lambda_i (x_i - x_l),$$

where $l \neq j$, which is in turn equivalent to the fact that

$$\sum_{i=1}^{k} \lambda_i \begin{pmatrix} x_i \\ 1 \end{pmatrix} = 0 \tag{2.5}$$

has a solution with $\lambda_j \neq 0$ (to see this, let $\lambda_l = 1 - \sum_{i \neq l, j} \lambda_i$ and $\lambda_j = -1$). Hence, $X$ is affinely independent if and only if (2.5) has the unique solution $\lambda_1 = \cdots = \lambda_k = 0$.

### 2.1.3 Polyhedra

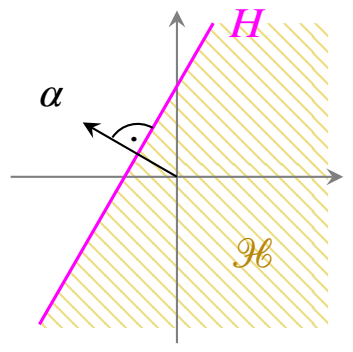

Figure 2.5: A hyperplane $H$ and corresponding halfspace $\mathscr{H}$.

A *polyhedron* is, intuitively speaking, a closed convex body whose surface decomposes into "flat" pieces. Mathematically, this flatness is grasped by the concept of *halfspaces*.

**2.4 Definition (polyhedra and polytopes):** Let $n \in \mathbb{N}$. A subset $H \subseteq \mathbb{R}^n$ is called a *hyperplane* of $\mathbb{R}^n$ if there exist $a \in \mathbb{R}^n$, $a \neq 0$ and $\beta \in \mathbb{R}$ such that

$$H = \left\{ x \in \mathbb{R}^n \colon a^T x = \beta \right\}.$$

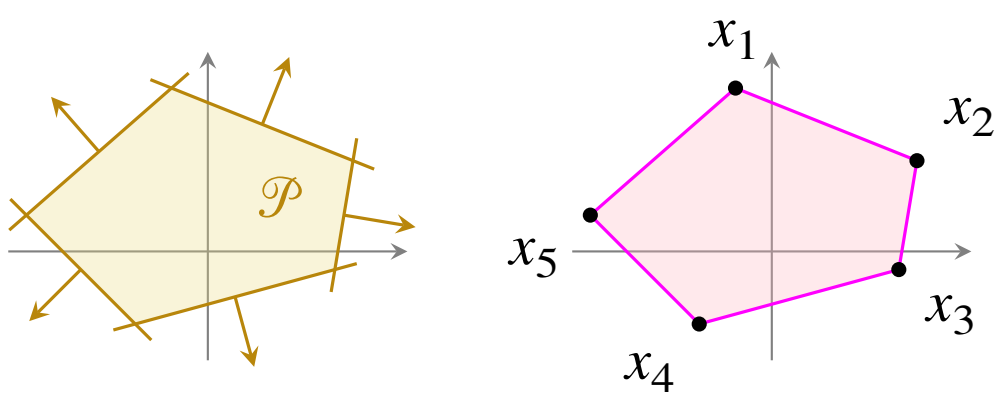

Figure 2.6: Example of a 2-dimensional polytope (its affine hull is $\mathbb{R}^2$) defined by five halfspaces (left) or equivalently as the convex hull of its five vertices $x_1, \dots, x_5$ (right).

Likewise, a (closed) *halfspace* of $\mathbb{R}^n$ is a set of the form

$$\mathcal{H} = \left\{x \in \mathbb{R}^n : a^T x \leq \beta\right\},$$

where again $0 \neq a \in \mathbb{R}^n$ and $\beta \in \mathbb{R}$. One says in the above situation that the hyperplane or halfspace, respectively, is *induced by* the pair $(a, \beta)$. The intersection of finitely many halfspaces is called a *polyhedron*. A *polytope* is a polyhedron that is bounded.

The *dimension* $\dim(\mathcal{P})$ of a polyhedron $\mathcal{P}$ is defined to be the dimension of $\operatorname{aff}(\mathcal{P})$, if $\mathcal{P} \neq \emptyset$, and $-1$ otherwise. In both cases $\dim(\mathcal{P})$ is one less than the maximum number of affinely independent vectors in $\mathcal{P}$. ◁

Polyhedra are the fundamental structure of linear and integer linear optimization. From the above definition, it follows that for a polyhedron $\mathcal{P}$ there is a matrix $A \in \mathbb{R}^{m \times n}$ and a vector $b \in \mathbb{R}^m$, for some $m \in \mathbb{N}$, such that

$$\mathcal{P} = \mathcal{P}(A, b) = \{x \in \mathbb{R}^n : Ax \leq b\},$$

i.e., $\mathcal{P}$ is the solution set of a system of linear inequalities, each of which defines a halfspace.

Note that polyhedra, being intersections of convex sets, are convex themselves. Complementary to the above *implicit* definition as the solution set of a system $Ax \leq b$, every polyhedron admits, by a theorem of Minkowski, an *explicit* characterization by means of convex and conic combinations.

**2.5 Theorem (Minkowski):** *The set $\mathcal{P} \subseteq \mathbb{R}^n$ is a polyhedron if and only if there are finite sets $V, W \subseteq \mathbb{R}^n$ such that $\mathcal{P} = \operatorname{conv}(V) + \operatorname{conic}(W)$.* ◁

If $\mathcal{P}$ in Theorem 2.5 is a polytope, then, since every nonempty cone is unbounded, $W = \emptyset$ must hold; hence, every polytope is the convex hull of its so-called *vertices* or *extreme points*, which are the "corners" of the polytope as shown in Figure 2.6. Vertices are a special instance of *faces* of a polyhedron, which are defined next.

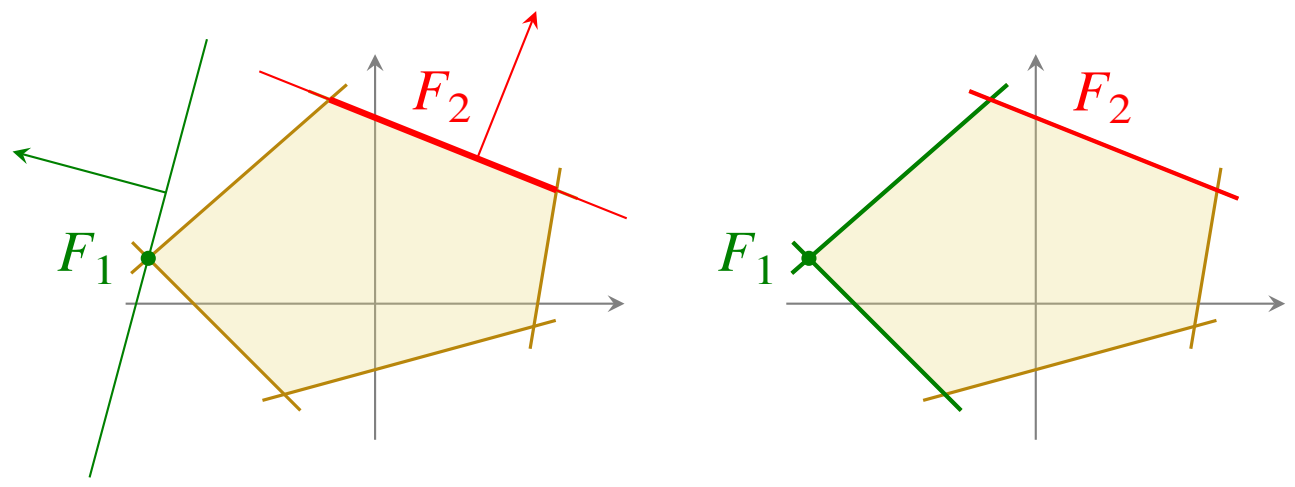

Figure 2.7: Two faces of a polytope, induced either by valid inequalities with $\dim(F_1) = 0$ and $\dim(F_2) = 1$ (left) or by their defining equality sets (highlighted on the right).

**2.6 Definition:** Let $\mathscr{P} \subseteq \mathbb{R}^n$ be a polyhedron. An inequality of the form

$$a^T x \leq \beta \tag{2.6}$$

with $a \in \mathbb{R}^n$ and $\beta \in \mathbb{R}$ is said to be *valid* for $\mathscr{P}$ if it is satisfied for all $x \in \mathscr{P}$. In that event, the set

$$F_{a,\beta} = \left\{x \in \mathscr{P} : a^T x = \beta\right\}$$

constitutes a *face* of $\mathscr{P}$, namely the *face induced by* (2.6). A zero-dimensional face is called a *vertex* of $\mathscr{P}$, one-dimensional faces are *edges* of $\mathscr{P}$. A face $F$ for which $\dim(F) = \dim(\mathscr{P}) - 1$ is called a *facet* of $\mathscr{P}$. ◁

Note that a face of a polyhedron is a polyhedron itself, as it is obtained by adding the inequalities $\alpha^T x \leq \beta$ and $\alpha^T x \geq \beta$ to the system $Ax \leq b$.

For a representation $\mathscr{P}(A, b)$ of the polytope $\mathscr{P}$, each face has another characterization (see Figure 2.7 for an example).

**2.7 Lemma:** *Let $\mathscr{P} = \mathscr{P}(A, b)$ with $A \in \mathbb{R}^{m \times n}$. For $E \subseteq \{1, \dots, m\}$, the set*

$$F_E = \left\{x \in \mathscr{P} : A_{E,\bullet} x = b_E\right\}$$

*is a face of $\mathscr{P}$. If $F_E \neq \emptyset$, its dimension is $n - \operatorname{rank}(A_{\operatorname{eq}(F_E),\bullet})$, where $\operatorname{eq}(F_E)$ is the* equality set *of $F_E$ defined by $\operatorname{eq}(F_E) = \left\{i : A_{i,\bullet} x = b \text{ for all } x \in F_E\right\}$.* ◁

Facets are of special importance because they are necessary and sufficient to describe a polyhedron:

**2.8 Theorem:** *Let $\mathscr{P} = \mathscr{P}(A, b)$ be a polyhedron, and assume that no inequality in $Ax \leq b$ is redundant, i.e., could be removed without altering $\mathscr{P}$. Let $I \cup J$ denote the partition of row indices defined by $I = \left\{i : A_{i,\bullet} x = b_i \text{ for all } x \in \mathscr{P}\right\}$. Then, the inequalities in $A_{J,\bullet} x \leq b_J$ are in one-to-one correspondence (via Lemma 2.7) to the facets of $\mathscr{P}$.* ◁

To describe a polyhedron $\mathcal{P}$, we thus need only $n - \dim(\mathcal{P})$ equations plus as many inequalities as $\mathcal{P}$ has facets. Any inequality that induces neither a facet nor the whole polytope $\mathcal{P}$ can be dropped without changing the feasible set, and every system $\tilde{A}x \leq \tilde{b}$ describing $\mathcal{P}$ needs to include at least one facet-inducing inequality for every facet of $\mathcal{P}$.

## 2.2 Linear Programming

A *linear program (LP)* is an optimization problem that asks for the minimization of a linear functional over a polyhedron. The most simple form would be

$$\min \quad c^T x \tag{2.7a}$$
$$\text{s.t.} \quad Ax \leq b, \tag{2.7b}$$

while an LP is said to be in *standard form* if it is stated as follows:

$$\min \quad c^T x \tag{2.8a}$$
$$\text{s.t.} \quad Ax = b \tag{2.8b}$$
$$x \geq 0. \tag{2.8c}$$

In both cases, $x \in \mathbb{R}^n$, $A \in \mathbb{R}^{m \times n}$ and $b \in \mathbb{R}^m$ are given, and we denote the feasible set by the letter $\mathcal{P}$ (for (2.7), $\mathcal{P} = \mathcal{P}(A, b)$ in the notation of Section 2.1). Note that both forms are equivalent in the sense that each can be transformed into the other. For example, given an LP in polyhedral form, we can replace $x$ by variables $x^+ \in \mathbb{R}^n$ and $x^- \in \mathbb{R}^n$, representing the positive and negative part of $x$, respectively, and introduce auxiliary variables $s \in \mathbb{R}^m$ to rewrite (2.7) in standard form as

$$\begin{aligned} \min \quad & c^T x^+ - c^T x^- \\ \text{s.t.} \quad & Ax^+ - Ax^- + s = b \\ & x^+ \geq 0,\ x^- \geq 0,\ s \geq 0. \end{aligned}$$

Moreover, if we had a *maximization* problem with objective $\max\ c^T x$, it could be converted to the above forms by the relation

$$\max \left\{ c^T x \colon x \in \mathcal{P} \right\} = -\min \left\{ -c^T x \colon x \in \mathcal{P} \right\}.$$

In view of this equivalence of different LP forms, we will in the following assume whatever form allows for a clear presentation.

Note that if (2.7) has an optimal solution $x^*$ with objective value $z^* = c^T x^*$, we can represent the set of optimal solutions by $\left\{ x \colon Ax \leq b \text{ and } c^T x = z^* \right\}$. This shows that the optimal set is always a *face* of $\mathcal{P}$. In addition, it is easy to show that if $\mathcal{P}$ has *any* vertex (which is always the case if the LP is in standard form), then *every* nonempty face of $\mathcal{P}$ contains a vertex. Hence we can conclude:

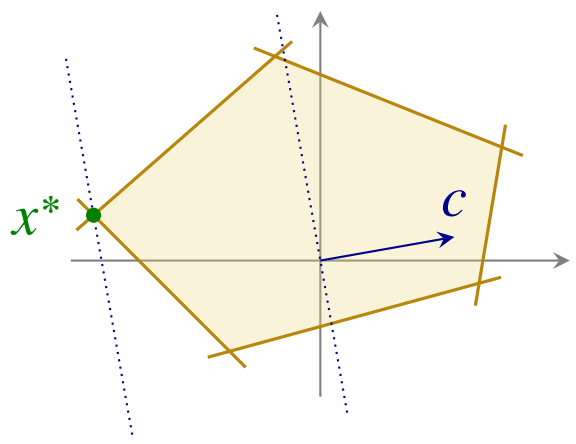

Figure 2.8: Example objective $c$ such that $c^T x$ is minimized for $x^*$. The two dotted lines show the hyperplanes $c^T x = 0$ and $c^T x = c^T x^*$, respectively.

**2.9 Observation:** *If an LP in standard form has a finite optimal objective value, there is always a vertex of $\mathscr{P}$ which is an optimal solution of the LP (see Figure 2.8).* ◁

The rest of this section is about characterizing and algorithmically finding such an optimal vertex.

### 2.2.1 Duality

One of the most important concepts in linear programming is that of *duality*, meaning that certain structures always occur in closely related *pairs*. In optimization, those structures include convex cones and systems of linear (in)equalities. For our purposes, the most important result is that of *LP duality*, a close relation of two LPs, as reviewed below.

**2.10 Definition:** Let an LP in standard form (2.8) be given; we call this the *primal* problem. The associated LP

$$\max \quad b^T y \tag{2.9a}$$

$$\text{s.t.} \quad A^T y \leq c \tag{2.9b}$$

is called the linear-programming *dual* of (2.8). ◁

Since we have seen that any LP can be transformed into standard form, one can also compute a dual for every LP. In particular, it is easy to verify that the dual of the dual results in the primal again. The motivation for LP duality lies in the following fundamental theorem.

**2.11 Theorem (strong duality):** *Assume that either (2.8) or (2.9) are feasible. Then*

$$\min \left\{ c^T x \colon Ax = b,\ x \geq 0 \right\} = \max \left\{ b^T y \colon A^T y \leq c \right\},$$

*where we include the values $\pm\infty$ as described on page 13. If both are feasible, then both have an optimal solution.* ◁

Note that Theorem 2.11 implies the statement of *weak duality*, namely that whenever $x$ is feasible for the primal and $y$ is feasible for the dual, then $c^T x \geq b^T y$.

LP duality is extremely useful because it allows for very compact proofs of optimality: if one wants to show that a certain solution $x^*$ of the primal LP is optimal, it suffices to provide a dual feasible $y^*$ with the property that $c^T x^* = b^T y^*$. Such a $y^*$ is called a *witness* for the optimality of $x^*$.

### 2.2.2 Primal and Dual Basic Solutions

In this section, we show how to represent a vertex of $\mathscr{P}$ by means of a *basis*. It is assumed that an LP is given in standard form (2.8) and that $A$ has full row rank $m$.

By Lemma 2.7, any vertex $\bar{x}$ of $\mathscr{P}$, which is a 0-dimensional face, can be characterized by a subset of the constraints of (2.8) that is fulfilled with equality and has rank $n$. Since (2.8b) has rank $m$ by assumption, $\bar{x}$ is a vertex if and only if there is an index set $N$ with $|N| = n - m$ such that $\bar{x}$ is the unique solution of the system

$$Ax = b,\ x_N = 0. \tag{2.10}$$

For $i \in N$ we can thus disregard the corresponding $i$-th column of the system $Ax = b$. Hence, we can represent $\bar{x}$ by $m$ linearly independent columns of $A$. Such a submatrix of $A$ is called a *(simplex) basis* and the corresponding set of column indices is denoted by $B = \{1, \ldots, n\} \setminus N$. The variables $x_B$ are called the *basic variables*, $x_N$ are the *non-basic variables*. By (2.10), every vertex $\bar{x}$ is a *basic solution* for (2.8b), i.e.,

$$\bar{x}_B = A_{\bullet,B}^{-1} b \quad \text{and} \quad \bar{x}_N = 0 \tag{2.11}$$

for a basis $B$ of $A$. Conversely, an arbitrary basic solution of the form (2.11) is a vertex of $\mathscr{P}$ only if additionally $\bar{x} \geq 0$, i.e., it is feasible for (2.8), then called a *basic feasible solution (BFS)* of the LP. Concludingly, $\bar{x}$ being a BFS is necessary and sufficient for $\bar{x}$ being a vertex of $\mathscr{P}$, while it should be noted that, in general, more than one BFS may correspond to the same vertex. Figure 2.9 shows a BFS for an example LP.

Let $B$ be a basis of $A$ and denote the feasible region of the dual (2.9) by $\mathscr{D}$. Arguing similarly as above, one can show that a vertex $\bar{y}$ of $\mathscr{D}$ must fulfill $A_{\bullet,B}^T \bar{y}_B = c_B$ (read $A_{\bullet,B}^T$ as $(A_{\bullet,B})^T$) and, in order to be feasible, also

$$c - A^T y \geq 0 \tag{2.12}$$

needs to hold. Hence the vector $\bar{y}$ defined by $\bar{y}^T = c_B^T A_{\bullet,B}^{-1}$ is called the *dual basic solution* associated to $\bar{x}$ defined in (2.11); it is a *dual BFS* if $\bar{y} \in \mathscr{D}$, i.e., if (2.12) holds.

$$\begin{pmatrix} \frac{1}{2} & 1 \end{pmatrix} \begin{pmatrix} x_1 \\ x_2 \end{pmatrix} = 1$$
$$x_1,\ x_2 \geq 0$$

(a) Definition of the LP.

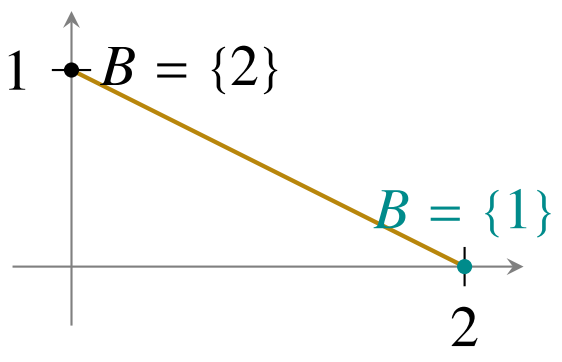

(b) Feasible space of the LP in $\mathbb{R}^2$.

Figure 2.9: Example of a linear program. For the basis $B = \{1\}$, we have $A_{\bullet,B}^{-1} b = 2 \cdot 1 = 2$, so the corresponding BFS is $\bar{x} = (\bar{x}_B, \bar{x}_N) = (2, 0)$.

### 2.2.3 The Simplex Method

If $\bar{x}$ and $\bar{y}$ are an associated pair of primal and dual basic solutions for a basis $B$ of $A$, it holds that

$$c^T \bar{x} = c_N^T \bar{x}_N + c_B^T \bar{x}_B = 0 + c_B^T A_{\bullet,B}^{-1} b = b^T \bar{y},$$

i.e., the objective values of $\bar{x}$ and $\bar{y}$ for the primal and dual LP, respectively, coincide. In view of Observation 2.9 and Theorem 2.11 this shows that solving an LP is tantamount to finding a basis $B$ for which the associated primal and dual basic solutions $\bar{x}$ and $\bar{y}$ are both feasible ($\bar{y}$ then is a witness of the optimality of $\bar{x}$). The several variants of the *simplex method* comprise algorithms that determine such a basis by a sequence of *basis exchange* operations in each of which a single element of $B$ is exchanged.

To be more specific, denoting the objective value by $z = c^T x$, by simple calculations starting from the form $A_{\bullet,B} x_B + A_{\bullet,N} x_N = b$ of (2.8b) we obtain the following representation

$$\begin{array}{rclclclcl} z & = & c_B^T A_{\bullet,B}^{-1} b & + & (c_N^T - c_B^T A_{\bullet,B}^{-1} A_{\bullet,N}) x_N & = & \bar{z} & + & \bar{c}_N^T x_N \\ x_B & = & A_{\bullet,B}^{-1} b & - & A_{\bullet,B}^{-1} A_{\bullet,N} x_N & = & \bar{b} & - & \bar{A}_N x_N \end{array} \tag{2.13}$$

of $z$ and $x_B$ with respect to $B$ in dependence of the values of $x_N$, where in the second step we have introduced suitable abbreviations $\bar{z}$, $\bar{b}$, $\bar{c}_N$ and $\bar{A}_N$. In this form, we can immediately read off the values $\bar{b}$ of the basic variables and the objective value $\bar{z}$ for the current basic solution that is defined by $x_N = 0$. The vector $\bar{c}_N^T = (c_N^T - \bar{y}^T A_N)$ encodes the dual feasibility (2.12) of that basis. Consequently $B$ must be an optimal basis if both $\bar{b} \geq 0$ (primal feasibility) and $\bar{c}_N \geq 0$ (dual feasibility) hold in (2.13).

Otherwise, we can perform a *simplex step*: assume that the $(i, k)$-th entry of $\bar{A}_N$ is non-zero. It can be shown that by performing a Gaussian *pivot* on that entry, i.e., turning the relevant column of (2.13) into a unit vector by elementary row operations, one essentially computes a representation of the form (2.13) with respect to the *adjacent basis* $B' = B \setminus \{i\} \cup \{j\}$, where $j \in N$ is the $k$-th entry of $N$. This notion of *adjacency* translates to the geometric interpretation, since vertices corresponding to adjacent basises always share an edge of the polyhedron.

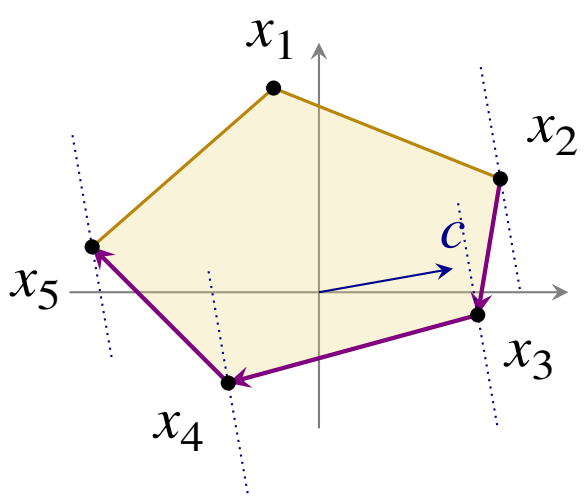

Figure 2.10: An example execution of the primal simplex algorithm, starting in $x_2$ and performing three basis exchanges until the optimal $x_5$ is reached. Along the path, the objective function $c^T x_i$ (shown by the blue dotted lines) is decreasing.

The *primal simplex algorithm* starts with a primal BFS (consult the literature for a method called *phase 1* to find such an initial BFS) and then iteratively performs the following steps:

(1) Choose a column (variable entering the basis) for which $\bar{c}_N$ is negative, i.e., the corresponding entry of $\bar{y}$ not yet dually feasible. This ensures that $z$ is nonincreasing, and it usually decreases.

(2) Choose a row (variable leaving the basis) in such a way as to ensure that the subsequent simplex step maintains primal feasibility; this can be achieved by a simple test called *min-ratio rule.*

(3) Perform the simplex step by pivoting on the column and row selected above.

The corresponding sequence of objective function values is nonincreasing. Under simple conditions on the method of selecting indices, one can show that this procedure results in an optimal basis, indicated by $\bar{c}_N \geq 0$, after a finite number of steps. See Figure 2.10 for an informal example.

The *dual simplex algorithm*, as the name suggests, sets off from a dual BFS ($\bar{c}_N \geq 0$) and then does essentially the same as its primal counterpart (with the role of rows and columns of (2.13) swapped during the basis exchange), maintaining dual feasibility and a nondecreasing objective function until primal feasibility ($\bar{b} \geq 0$) is established.

Numerous variants and optimizations of the basic method described above exist. An important one is the so-called *revised simplex* which is based on the observation that, especially for $n \gg m$, it is wasteful to pivot the complete system (2.13) in each step. Instead, one maintains a representation of $A_{\bullet,B}^{-1}$ (usually in the form of an *LU factorization*), which can be shown to be sufficient to carry out an iteration of the algorithm. Furthermore, it should be noted that there exists an efficient method to incorporate *upper bounds* on the variables, e.g. of the form $0 \leq x \leq 1$, without having to increase the size of the formulation by the explicit addition of constraints $x \leq 1$.

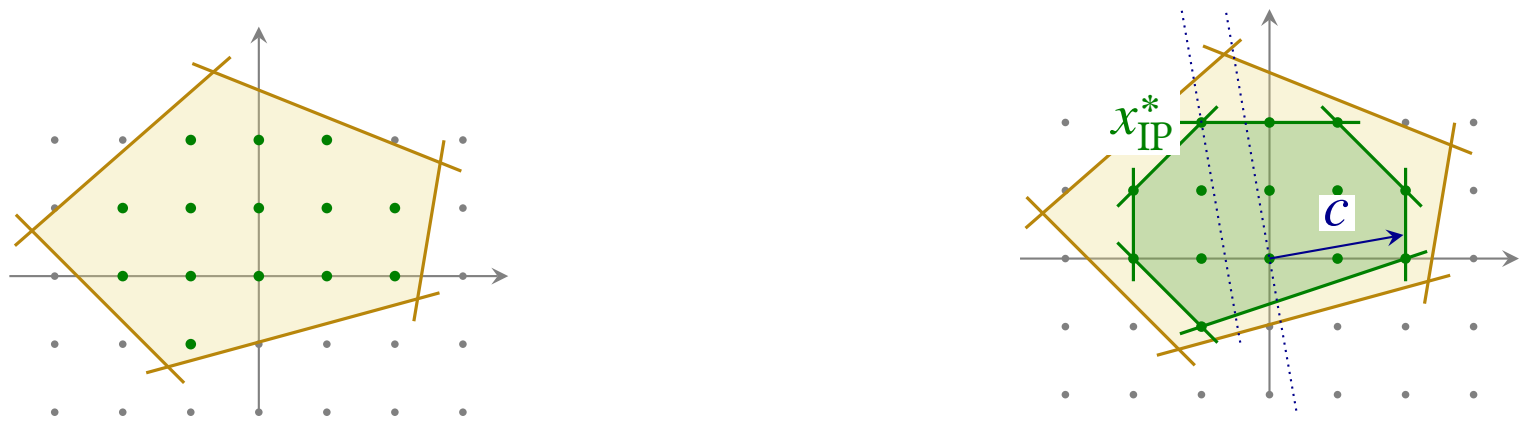

(a) Feasible set $\mathscr{P} \cap \mathbb{Z}^2$ and LP relaxation.

(b) Equivalent polytope $\mathscr{P}_{\mathrm{I}} = \mathrm{conv}(\mathscr{P} \cap \mathbb{Z}^2)$ and optimal IP solution $x^*_{\mathrm{IP}}$.

Figure 2.11: Feasible space of an integer program (IP) and its LP relaxation.

It has been shown that the worst-case complexity of the simplex algorithm is exponential in the problem size [KM72]. The very contrary *empirical* observation however is that the number of pivots before optimality is usually in $O(m)$. This explains why, although LP solving algorithms with polynomial worst-case complexity exist, the simplex method is still the most prevalent one in practice.

## 2.3 Integer Programming

In integer programming, we are concerned with LPs augmented by the additional requirement that the solution be *integral*. Formally, we define an *integer linear program (IP)* as an optimization problem of the form

$$\begin{aligned} \min \quad & c^T x && \text{(2.14a)} \\ \text{s.t.} \quad & Ax \le b && \text{(2.14b)} \\ & x \in \mathbb{Z}^n, && \text{(2.14c)} \end{aligned}$$

where we assume that all entries of $A$, $b$ and $c$ are rational. The LP that results when replacing (2.14c) by $x \in \mathbb{R}^n$ is called its *LP relaxation* (see Figure 2.11(a)). Let $\mathscr{P} = \mathscr{P}(A, b)$ as before denote the feasible set of the LP relaxation and $\mathscr{P}_{\mathrm{I}} = \mathrm{conv}\,(\mathscr{P}(A, b) \cap \mathbb{Z}^n)$ the convex hull of integer points in $\mathscr{P}$. Under the above assumption, it can be shown that $\mathscr{P}_{\mathrm{I}}$ is a polyhedron. Note that solving (2.14) is essentially equivalent to solving $\min\{c^T x \colon x \in \mathscr{P}_{\mathrm{I}}\}$. Thus an IP can, in principle, be solved by an *LP*: if $A'$ and $b'$ are such that $\mathscr{P}_{\mathrm{I}} = \mathscr{P}(A', b')$, then the optimal IP solution is also optimal for the LP

$$\begin{aligned} \min \quad & c^T x \\ \text{s.t.} \quad & A' x \le b' \\ & x \in \mathbb{R}^n; \end{aligned}$$

see Figure 2.11(b) for an example.

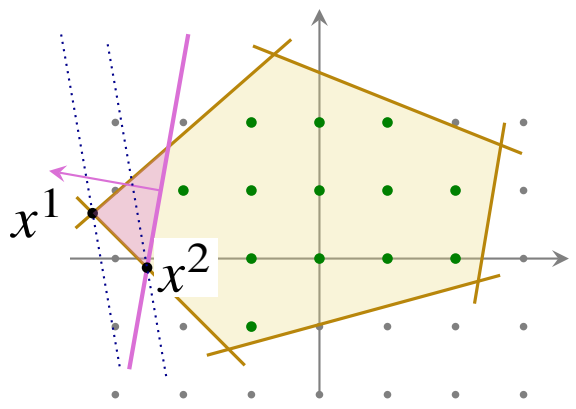

Figure 2.12: Example cutting plane that cuts the non-integral initial LP solution $x^1$ but is valid for $\mathscr{P}_\mathrm{I}$; it implies a new LP solution $x^2$ with improved objective value.

The problem is that, in general, it is hard to derive a description of $\mathscr{P}_\mathrm{I}$ from $\mathscr{P}$. In fact, IPs are NP-hard to solve in general, whereas we have seen above that linear programming is contained in P.

Two complementary approaches for solving IPs are important in our context: one is to tighten the LP relaxation (2.14) by adding *cuts*, i.e., inequalities that are valid for $\mathscr{P}_\mathrm{I}$ but not for $\mathscr{P}$. The other, named *branch-&-bound*, is about recursively dividing the feasible space $\mathscr{P} \cap \mathbb{Z}^n$ into smaller subproblems among which the optimal solution is searched, interleaved with the generation of bounds that allow to skip most of these subproblems.

### 2.3.1 Cutting Planes

Assume that we try to solve the IP (2.14) by solving its LP relaxation, i.e., minimize $c^T x$ over $\mathscr{P}$ instead of $\mathscr{P}_\mathrm{I}$. If the LP solution $x^1$ happens to be integral, it is clear that $x^1$ is also optimal for (2.14). Otherwise, by Theorem 2.8 there must exist at least one inequality that is valid for $\mathscr{P}_\mathrm{I}$ but not for $x^1$. Any such inequality is called a *cutting plane* or simply *cut* for $x^1$ (Figure 2.12). If we add a cut to the LP relaxation (2.14) and solve the LP again, we necessarily get a new solution $x^2 \neq x^1$ (because the cut is violated by $x^1$). Since the feasible space was reduced, $c^T x^2 \geq c^T x^1$ must hold.

This method can be iterated as long as new cuts for the current solution $x^i$ can be generated, leading to a sequence of LP solutions $(x^i)_i$ such that $(c^T x^i)_i$ is monotonically increasing. If the cuts are "good enough" (especially if they include the facets of $\mathscr{P}_\mathrm{I}$), some $x^k$ will eventually be feasible for $\mathscr{P}_\mathrm{I}$ and hence equal the IP solution $x^*$. While there exists a cut-generation algorithm, called *Gomory-Chvátal method*, that provably terminates in $x^*$ after a finite number of steps, the number of cuts it usually introduces is prohibitive for practical applications. For many classes of IPs, however, there exist special methods to derive cuts that are based on the specific structure of the problem—we will encounter such a case in Section 4.3.

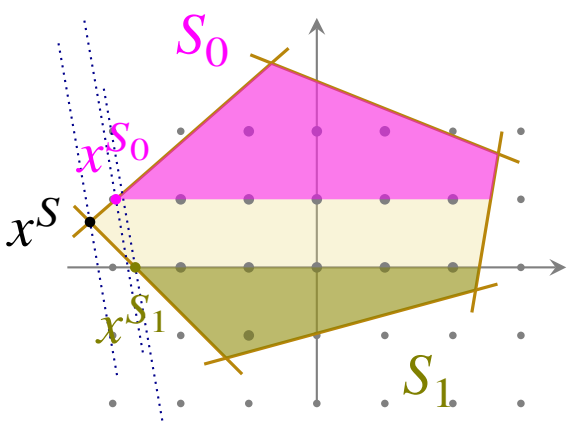

Figure 2.13: The branching principle: All integral points of $\mathcal{P}$ are contained in either $S_0 = \{x \in \mathcal{P} : x_2 \geq 1\}$ or $S_1 = \{x \in \mathcal{P} : x_2 \leq 0\}$, and both solutions $x^{S_0}$ and $x^{S_1}$ have larger objective value than $x^S$.

### 2.3.2 Branch-&-Bound

Let us again assume that we solve the LP relaxation of (2.14) and obtain a non-integral LP solution $x^\emptyset \in \mathcal{P} \setminus \mathbb{Z}^n$ with, say, $x_i^\emptyset \notin \mathbb{Z}$. The key idea of LP-based *branch-&-bound* is to define two disjoint subproblems $S_0$ and $S_1$ of the original problem $S_\emptyset = (2.14)$ with the property that the optimal IP solution $x^*$ of $S_\emptyset$ is contained in either $S_0$ or $S_1$. To be specific, note that necessarily $x_i^* \in \mathbb{Z}$, so either $x_i^* \leq \lfloor x_i^\emptyset \rfloor$ or $x_i^* \geq \lceil x_i^\emptyset \rceil$ needs to hold, which gives rise to the two subproblems

$$S_0: \quad \begin{aligned} \min \quad & c^T x \\ \text{s.t.} \quad & Ax \leq b \\ & x_i \leq \lfloor x_i^\emptyset \rfloor \\ & x \in \mathbb{Z}^n \end{aligned} \qquad \text{and} \qquad S_1: \quad \begin{aligned} \min \quad & c^T x \\ \text{s.t.} \quad & Ax \leq b \\ & x_i \geq \lceil x_i^\emptyset \rceil \\ & x \in \mathbb{Z}^n. \end{aligned}$$

For both, we can again solve the LP relaxation (with the additional constraint on $x_i$) to obtain LP solutions $x^0$ and $x^1$ with objective function values $z^0$ and $z^1$, respectively (cf. Figure 2.13). If both solutions are integral, clearly the one with smaller objective function value is optimal for (2.14). Otherwise, the process can be recursed to split a subproblem into two sub-subproblems (e.g., $S_{00}$ and $S_{01}$), and so forth, creating a binary tree of “problem nodes” whose leaves accord to problems that either have an integral LP solution or are infeasible—in both cases, no further subdivision is necessary.

While this technique already reduces the search space in a substantial way, using advanced *bounding* allows to further reduce the size of the abovementioned tree. Note that any *feasible* solution $\hat{x} \in \mathcal{P} \cap \mathbb{Z}^n$ gives an upper bound on the optimal objective value $z^* = c^T x^*$, i.e., $c^T \hat{x} \geq c^T x^*$. Furthermore, in any subproblem $S$, the objective value $z^S$ of the optimal solution $x^S$ to the corresponding LP relaxation is a *lower bound* on the optimal integral solution value $\hat{z}^S$ *of that subproblem*. If we now solve the LP relaxation of some subproblem $S$ obtaining $z^S$ and it holds that $z^S \geq c^T \hat{x}$ for any feasible $\hat{x}$ that has been found earlier, there is no need to subdivide $S$ (even if $x^S$ is not integral): no feasible point of $S$ can improve upon the objective value of $\hat{x}$.

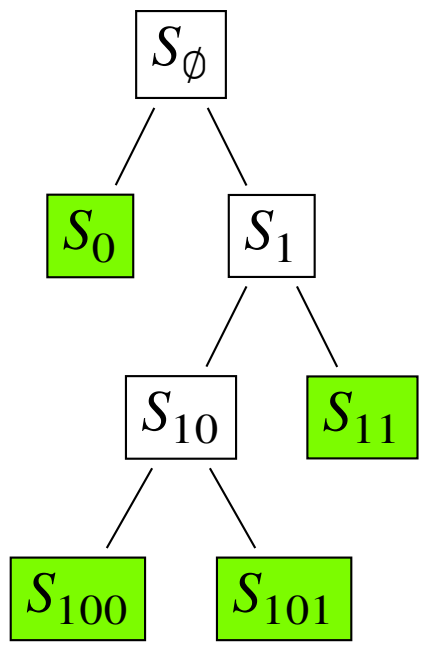

Figure 2.14: An example branch-and-bound enumeration tree; the leaves are not further explored because the respective LPs are either infeasible or have integral solutions.

If we otherwise split $S$ into subproblems $S'$ and $S''$ and solve the respective LP-relaxations to obtain $z^{S'}$ and $z^{S''}$, we can conclude that $\min\{z^{S'}, z^{S''}\} \leq \hat{z}^S = c^T\hat{x}^S$, i.e., the smaller of both is a lower bound on $\hat{z}^S$, because the optimal integral solution $x^S$ for $S$ must be in either $S'$ or $S''$. If this minimum is larger than $z^S$, we can *improve* the lower bound for $S$. This bound update can possibly be propagated to the parent of $S$ if $S$ has a sibling for which a lower bound has already been computed, and so forth.

The algorithm terminates if there are either no unexplored subproblems left, or if a feasible solution $\hat{x}$ for (2.14) and a lower bound $z^\emptyset$ on the optimal objective value $z^*$ has been found such that $\hat{x} = z^\emptyset$: it is then clear that no better solution than $\hat{x}$ exists, so $\hat{x}$ must be optimal. An example of the resulting search tree is shown in Figure 2.14.

In practice, the cutting-plane and branch-&-bound approach are often interleaved within a *branch-&-cut* algorithm, where cutting planes might be inserted in each node of the branch-&-bound tree. See e.g. [Hel+14] for an application to ML decoding.

## 2.4 Combinatorial Optimization

An optimization problem is called *combinatorial* if the feasible set comprises all subsets of a finite ground set $\Xi = \{\xi_1, \dots, \xi_n\}$ that fulfill a certain property, and the objective value of an $S \subseteq \Xi$ has the form $f(S) = \sum_{\xi \in S} c_\xi$ for given cost values $c_\xi \in \mathbb{R}$ associated to each element $\xi \in \Xi$.

A popular example is the *shortest-path problem*: given a graph $G = (V, E)$, two vertices $s, t \in V$ and cost $c_e$ associated to each edge $e \in E$, it asks for an $s$–$t$ path $P^*$ (in fact, only paths in which each edge occurs at most once are allowed) with minimum total cost $f(P^*)$, where the objective function $f(P) = \sum_{e \in P} c_e$ accumulates, for each edge $e$ visited by the path $P$, the cost value $c_e$ assigned to $e$. Here, the ground set consists of the set of edges $\Xi = E$, and a subset of edges is feasible if it forms (in appropriate ordering) a path in $G$. A small example is shown in Figure 2.15.

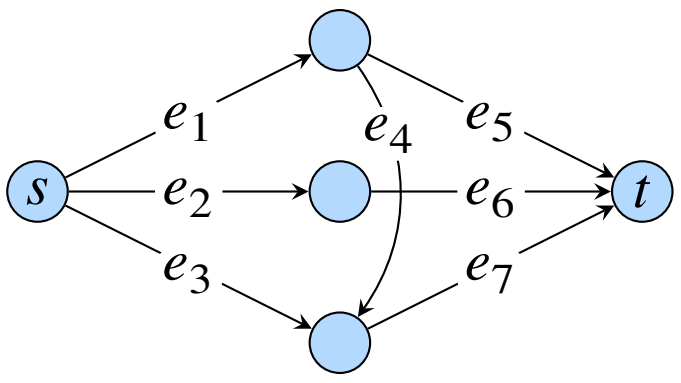

Figure 2.15: Example instance of the shortest-path problem, with the ground set $\Xi = \{e_1, \dots, e_7\}$ and feasible subsets ($s$-$t$-paths) $\{e_1, e_5\}$, $\{e_1, e_4, e_7\}$, $\{e_2, e_6\}$, and $\{e_3, e_7\}$.

In a combinatorial optimization problem, one can identify each subset $S \subseteq \Xi$ by its characteristic (or incidence) vector $x^S \in \{0,1\}^n$, where

$$x_i^S = \begin{cases} 1 & \text{if } \xi_i \in S, \\ 0 & \text{otherwise.} \end{cases}$$

Then, the set $X = \{x^S : S \subseteq \Xi \text{ feasible}\}$ that represents the feasible solutions is a subset of $\{0,1\}^n$. As furthermore $f(S) = c^T x^S$ with $c = (c_{\xi_1}, \dots, c_{\xi_n})^T$, every combinatorial optimization problem with such an objective function can be represented by the LP

$$\begin{aligned} \min \quad & c^T x \\ \text{s.t.} \quad & x \in \operatorname{conv}(X) \end{aligned}$$

whose feasible polytope $\mathscr{P}$ is a subset of the unit hypercube $[0,1]^n$.

In case of the shortest $s$–$t$ path problem, an explicit formulation of the *path polytope*, i.e., the convex hull of incidence vectors of $s$–$t$ paths, is known. The corresponding LP to solve the shortest path problem is

$$\begin{aligned} \min \quad & c^T x = \sum_{e \in E} c_e x_e && && \text{(2.15a)} \\ \text{s.t.} \quad & \sum_{e \in \delta^+(s)} x_e - \sum_{e \in \delta^-(s)} x_e = 1 && && \text{(2.15b)} \\ & \sum_{e \in \delta^+(t)} x_e - \sum_{e \in \delta^-(t)} x_e = -1 && && \text{(2.15c)} \\ & \sum_{e \in \delta^+(v)} x_e - \sum_{e \in \delta^-(v)} x_e = 0 && \text{for all } v \notin \{s,t\} && \text{(2.15d)} \\ & x \in [0,1]^{|E|}, && && \text{(2.15e)} \end{aligned}$$

where (2.15b) and (2.15c) ensure that the path starts in $s$ and ends in $t$, respectively, and the so-called *flow conservation constraints* (2.15d) state that the path must leave any other vertex $v$ as often as it enters $v$.

In general, however, it is highly nontrivial to find an explicit representation of $\mathcal{P}$. For a large class of hard problems one can nevertheless at least formulate some LP-relaxation $\mathcal{P}' \supseteq \operatorname{conv}(X)$ such that $\mathcal{P}' \cap \{0,1\}^n = X$, i.e., there is an *integer* programming model

$$\begin{aligned} \min \quad & c^T x \\ \text{s.t.} \quad & x \in \mathcal{P}' \\ & x \in \{0,1\}^n \end{aligned}$$

of the problem, which can then be tackled by the methods presented in Section 2.3. For several problems—in this text, most notably the *decoding* problem introduced in the following chapter—this polyhedral approach to combinatorial optimization has led to the most efficient algorithms known.

# 3 Coding Theory Background

How can information be transmitted *reliably* via an *error-prone* transmission system? The mathematical study of this question initiated the emergence of *information theory* and, since one particularly important answer lies in the use of error-correcting codes, of *coding theory* as a mathematical discipline of its own right.

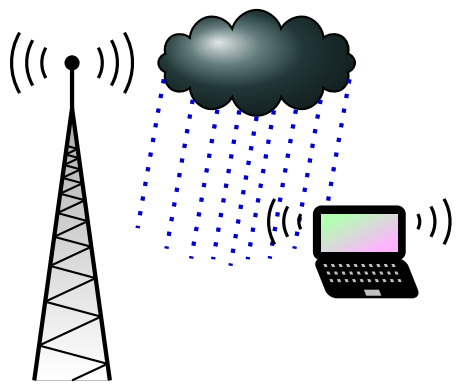

a noisy data transmission scenario

In contrast to *physical* approaches to increase the reliability of communication (e.g. increased transmit power, more sensitive antennas, ...), error-correcting codes offer a completely *ideal* solution to the problem. It has been shown that, by coding, an arbitrary level of reliability is achievable, despite the unavoidable inherent unreliability of any technological equipment—in fact, the unparalleled development of microelectronic devices and their ability to communicate via both wired and wireless networks would not have been possible without the accompanying progresses in coding theory.

This chapter reviews the basics of error-correcting codes and their application to reliable communication. For a more complete coverage of these topics, we refer the interested reader to the broad literature on the subject: the birth of information theory was Shannon's seminal work "A mathematical theory of communication" [Sha48]. Recommendable textbooks on information theory and its applications are [Mac03; Gal68]. There are several books covering "classical" coding theory (this term is elucidated later), e.g. [MS77], while modern aspects of coding are collected in [RU08].

## 3.1 System Setup and the Noisy-Channel Coding Theorem

The principle of error-correction coding is to preventively include *redundancy* in the transferred messages, thus communicating *more* than just the actual information, with the goal of enabling the receiver to recover that information, even in the presence of noise on the transmission channel. The general system setup we consider is as depicted in Figure 3.1:

- the function by which these "bloated" messages are computed from the original ones is called the *encoder*;
- the *channel* introduces noise to the transmitted signal, i.e., at the receiver there is *uncertainty* about what was actually sent;

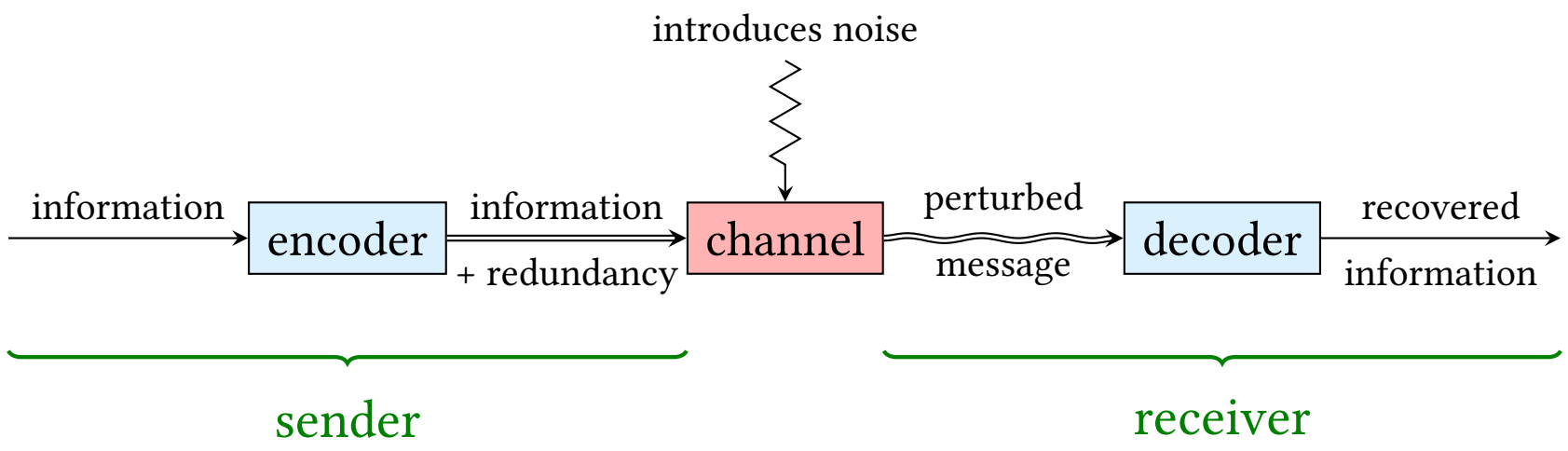

Figure 3.1: Model of the transmission system.

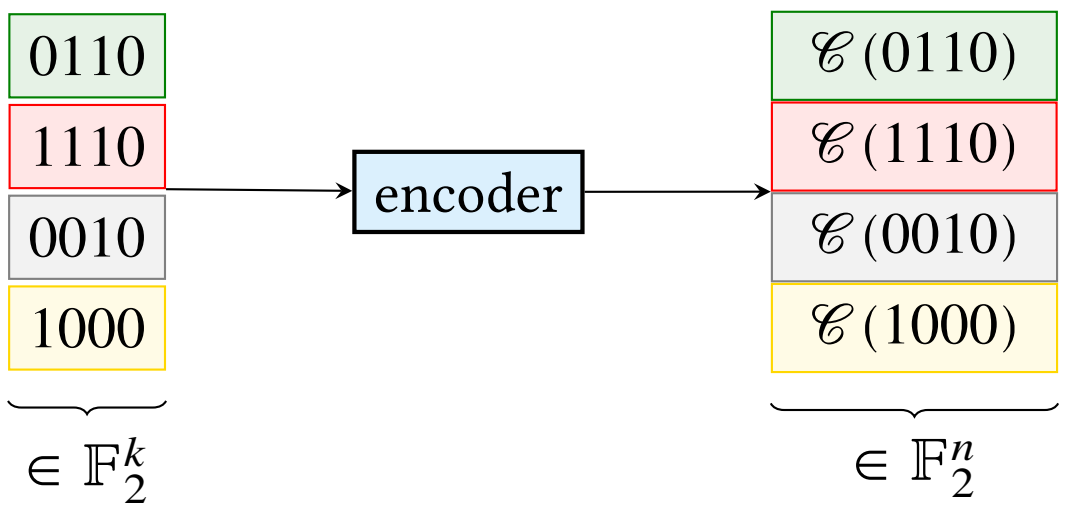

Figure 3.2: Block encoding of information words into codewords.

- the *decoder* tries to recover the original information from the received signal.

Throughout this text, we are concerned only with *block codes*, which means that the information enters the encoder in form of chuncks of uniform size (the *information words*), each of which is encoded into a unique coded message of again uniform (but larger) size (the *codewords*). For now, we additionally restrict ourselves to the *binary* case, i.e., the alphabet for both information and codewords is $\mathbb{F}_2 = \{0, 1\}$. This leads to the following definitions.

**3.1 Definition (code):** An $(n, k)$ *code* is a subset $\mathscr{C} \subseteq \mathbb{F}_2^n$ of cardinality $2^k$. A bijective map from $\mathbb{F}_2^k$ onto $\mathscr{C}$ is called an *encoding function for* $\mathscr{C}$.

Whenever the context specifies a concrete encoding function, and if there is no risk of ambiguity, the symbol $\mathscr{C}$ will be used interchangeably for both the code and its encoding function.

The numbers $k$ and $n$ are referred to as *information length* and *block length*, respectively. Their quotient $r = k/n < 1$ represents the amount of information per coded bit and is called the *rate* of $\mathscr{C}$. ◁

The resulting coding scheme is sketched in Figure 3.2.

The concept of redundancy is entailed by the fact that $\mathscr{C}$ is a strict (and usually very small) subset of the space $\mathbb{F}_2^n$, i.e., most of the vectors in $\mathbb{F}_2^n$ are *not* codewords, which is intended to make the codewords much easier to distinguish from each other than the informations words of $\mathbb{F}_2^k$.

Note that in this definition the encoder is secondary to the code. This reflects the fact that, for the topics covered by this text, the structure of the set of codewords is more important than the actual encoding function.

We assume that the channel through which the codewords are sent is *memoryless*, i.e., that the noise affects each individual bit independently; it thus can be defined as follows.

**3.2 Definition (binary-input memoryless channel):** A *binary-input memoryless channel* is characterized by an output domain $\mathscr{Y}$ and the two conditional probability functions

$$P(y_i \mid x_i = 0) \quad \text{and} \quad P(y_i \mid x_i = 1) \tag{3.1}$$

that specify how the output $y_i \in \mathscr{Y}$ depends on the two possible inputs 0 and 1, respectively (we assume that these two functions are not identical—otherwise, the output would be independent of the input and would thus not contain any information about the latter). Even more compactly, the frequently used *log-likelihood ratio (LLR)*

$$\lambda_i = \ln\left(\frac{P(y_i \mid x_i = 0)}{P(y_i \mid x_i = 1)}\right) \tag{3.2}$$

represents the entire information revealed by the channel about the sent symbol $x_i$. If $\lambda_i > 0$, then $x_i = 0$ is more likely than $x_i = 1$, and vice versa if $\lambda_i < 0$. The absolute value of $\lambda_i$ indicates the *reliability* of this tendency. ◁

An example of a channel transmission and resulting LLR values is sketched in Figure 3.3. When the receiver observes the result $\lambda \in \mathbb{R}^n$ (we mostly use LLRs in favor of $y \in \mathscr{Y}^n$ from now on) of the transmission of an encoded information word $x = \mathscr{C}(u)$ through the channel, it has to answer the following question: *which codeword $x \in \mathscr{C}$ do I believe has been sent, under consideration of $y$?* This "decision maker" is called the *decoder*, which is an algorithm realizing a decoding function

$$\textsc{decode}\colon \mathbb{R}^n \to \mathbb{R}^n; \tag{3.3}$$

the decoder is intentionally (for reasons that will become clear later) allowed to output not only elements of $\mathbb{F}_2^n$ but arbitrary points of the unit hypercube $[0,1]^n$ (which includes $\mathbb{F}_2^n$ via the canonical embedding). We speak of *decoding success* if $x \in \mathscr{C}$ was sent and $\textsc{decode}(\lambda) = x$, while a *decoding error* occurs if $\textsc{decode}(\lambda) = x' \neq x$, which includes the cases that $x' \in \mathscr{C}$, i.e., the decoder outputs a codeword but not the one that was sent, and $x' \notin \mathscr{C}$, i.e., the decoder does not output a codeword at all.

Assuming a *uniform prior* on the sender's side, i.e., that all possible information words occur with the same probability (*source coding*, the task of accomplishing this assumption, is not

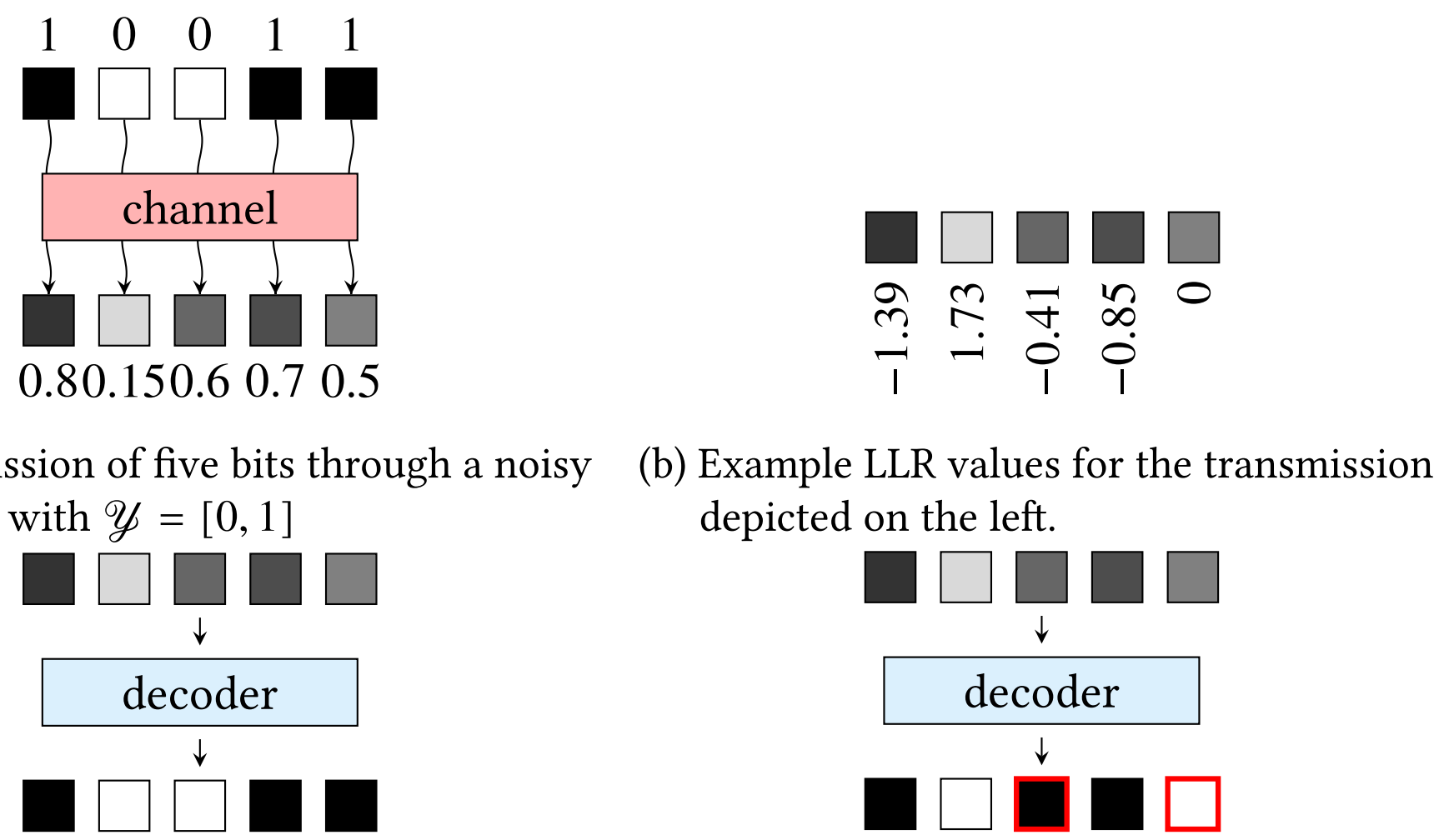

(a) Transmission of five bits through a noisy channel with $\mathcal{Y} = [0, 1]$

(b) Example LLR values for the transmission depicted on the left.

(c) Decoding success for the above transmission.

(d) Example of a decoding failure (two faulty bits) for the above transmission.

Figure 3.3: Sketch of transmission of five bits through a noisy channel with both successful (Figure 3.3(c)) and failed (Figure 3.3(d)) decoding.

covered here), the error-correcting performance of a communication system consisting of code, channel, and decoding algorithm can be evaluated by means of its average *frame-error rate*

$$\text{FER} = \frac{1}{|\mathscr{C}|} \sum_{x \in \mathscr{C}} P\left(\textsc{decode}(\lambda) = x \mid x \text{ was sent}\right). \tag{3.4}$$

We can now state the main task of coding theory: given a certain channel model (3.1), design an $(n, k)$ code and an accompanying decoder (3.3) such that the demands requested by the desired application are fulfilled, which may include:

- The frame-error rate (3.4) should be sufficiently small in order to ensure reliable communication.
- The rate $r = k/n$ should be as large as possible, because a small rate corresponds to a large number of transmitted bits per information bit, i.e., large coding overhead.
- The block length $n$ should be small: high block lengths generally increase the complexities of both encoder and decoder and may additionally introduce undesirable latencies in e.g. telephony applications.
- The complexity of the decoding algorithm needs to be appropriate.

It is intuitively clear that some of the above goals are opposed to each other. The first two, however, are not as incompatible as one might suspect—Claude Shannon proved a stunning result [Sha48] which implies that, at a *fixed* positive code rate, the error probability can be made arbitrarily small.

**3.3 Theorem (noisy-channel coding theorem):** *For given $\varepsilon > 0$ and $r < C$, where $C > 0$ depends only on the channel, there exists a code $\mathscr{C}$ with rate at least $r$, and a decoding algorithm for $\mathscr{C}$ such that the frame-error rate (3.4) of the system is below $\varepsilon$.* ◁

As beautiful as both the result and its proof (which is explained thoroughly in [Mac03]) are, they are unfortunately completely non-constructive in several ways:

- the decoding error probability vanishes only for the block length $n$ going to infinity;
- the proof makes use of a *random coding* argument, hence it does not say anything about the performance of a concrete, finite code;
- the running time of the theoretical decoding algorithm used in the proof is intractable for practical purposes.

As a consequence of the first two aspects, the search for and construction of "good" codes, i.e., codes that allow for the best error correction at a given finite block length $n$ and rate $r$, has emerged as an research area on itself, which is nowadays often nicknamed "classical coding theory." For a long time, however, the problem of decoder complexity was not a major focus of the coding theory community. The term "modern coding theory" nowadays refers to a certain paradigm shift that has taken place since the early 1990's, governed by the insight that *suboptimal codes* which are develeoped *jointly* with harmonizing low-complexity decoding algorithms can lead to a higher overall error-correcting performance in practical applications than the "best" codes, if no decoder is able to exploit their theoretical strength within reasonable running time (see [CF07] for the historical development of coding theory).

The rest of this chapter is organized as follows. In Section 3.2, we discuss both the optimal MAP and the ML decoding rule, which are equivalent in our case. Afterwards, the prevalent additive white Gaussian noise (AWGN) channel model is explained in Section 3.3. Section 3.4 introduces binary linear block codes, a subclass of general block codes that is most important in practice and with some exceptions assumed throughout this text. A special type of linear block codes, called turbo codes, is presented in Section 3.5. Finally, Section 3.6 explains how codes and channels can be generalized to the non-binary case.

Note that this chapter does not cover any specific *decoding* algorithm. The next chapter covers various decoding approaches using mathematical optimization, which comprises the major topic of this text. For other decoding algorithms, e.g. the ones that are used in today's electronic devices, we refer to the literature.

## 3.2 MAP and ML Decoding

An optimal decoder (with respect to frame-error rate) would always return the codeword that was sent *with highest probability* among all codewords $x \in \mathscr{C}$, given the observed channel output $y$:

$$x_{\mathrm{MAP}} = \arg\max_{x \in \mathscr{C}} P(x \mid y). \tag{3.5}$$

This is called *MAP decoding*. By Bayes' theorem, we have

$$P(x \mid y) = \frac{P(y \mid x) P(x)}{P(y)}.$$

Since $P(y)$ is independent of the sent codeword $x$ and by assumption $P(x)$ is constant on $\mathscr{C}$, we obtain the equivalent *ML decoding* rule:

$$x_{\mathrm{ML}} = \arg\max_{x \in \mathscr{C}} P(y \mid x). \tag{3.6}$$

Unfortunately, ML decoding is NP-hard in general [BMT78], which motivates the search for special classes of codes that are both strong and allow for an efficient decoding algorithm which at least approaches the ML error-correction performance. On the other hand, it is desirable to know the frame-error rate for a given code under *exact* ML decoding, because it *(a)* constitutes the ultimate theoretical performance measure of the code itself and *(b)* serves as a "benchmark" for the quality of suboptimal decoding algorithms.

## 3.3 The AWGN Channel

The most immediate and simple example of a binary-input memoryless symmetric channel as defined in Definition 3.2 is called the *binary symmetric channel (BSC)*: it flips a bit with probability $p < 1/2$ and correctly transmits it with probability $q = 1 - p$, and hence $\lambda_i = \pm \ln(p/q)$, i.e., there are only two possible channel outputs.

While the conceptual simplicity of the BSC is appealing, for practical applications it turns out to be simplified too much. Imagine a device in which some incoming electromagnetic signal is translated by a circuit (consisting of e.g. antenna, electronic filters etc.) into a voltage $v$ with expected values $v_0$ and $v_1$ for the transmission of a 0 and 1, respectively. For a BSC channel model, we could round $v$ to the closest of those values and pass that "hard" information (either 0 or 1) to the decoder. But clearly, knowing *how far* $v$ is from the value it is rounded to contains valuable information about the *reliability* of the received signal—a value of $v$ close to the mean $(v_1 - v_0)/2$ is less reliable than one close to either $v_0$ or $v_1$ (cf. the figures along Definition 3.2). Consequently, the decoder should take that "soft" information into account.

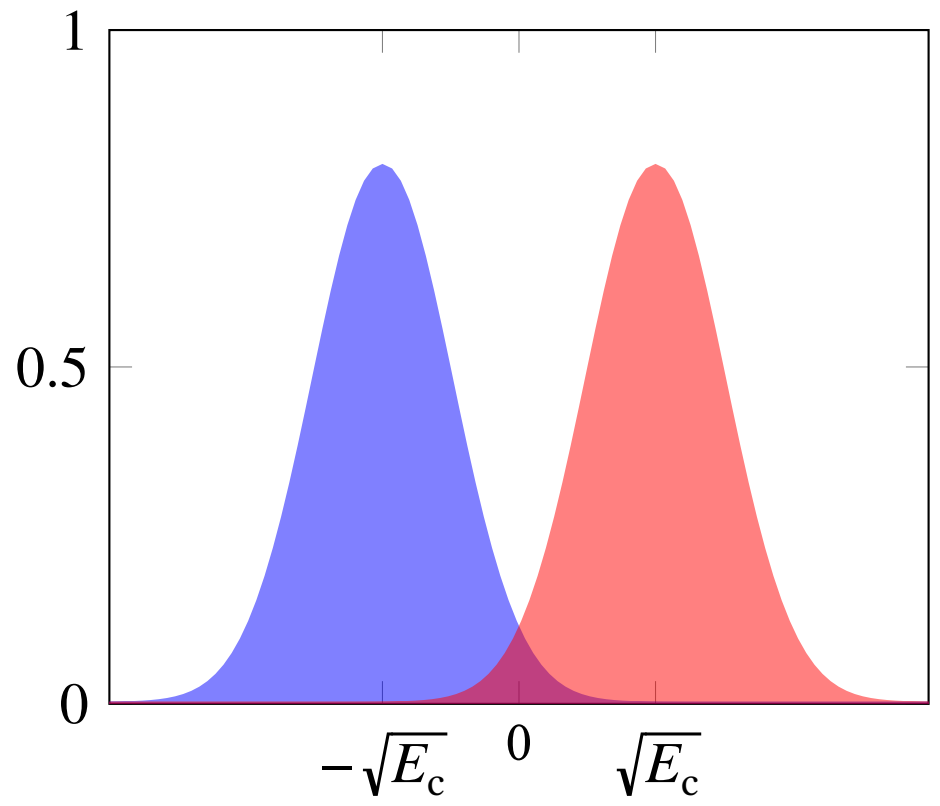

Figure 3.4: Probability density functions $P(y_i \mid x_i = 1)$ (left) and $P(y_i \mid x_i = 0)$ (right) of an AWGN channel with energy $E_c$.

The most prominent *soft-output* channel model is the *AWGN channel*, in which independent Gaussian noise (as it appears frequently in nature) is added to each transmitted symbol. It is characterized by a Gaussian distribution with mean $(-1)^{x_i}\sqrt{E_c}$ and variance $\sigma^2$, as shown in Figure 3.4. Here, $E_c$ is the transmit energy per channel symbol and $\sigma^2$ is the noise energy of the channel [RU08]:

$$P(y_i \mid x_i) = \frac{1}{\sqrt{2\pi\sigma^2}} e^{-\frac{1}{2}\cdot\left(\frac{y-(-1)^{x_i}\sqrt{E_c}}{\sigma}\right)^2}. \tag{3.7}$$

Note that the AWGN channel challenges the conceptual distinction between *transmission success* and *transmission error* in favor of an ubiquitous presence of noise: the expected value $\pm\sqrt{E_c}$ that corresponds to a “noiseless” transmission will be received only with probability zero.

For a given code rate $r$, an AWGN channel can be specified by a single quantity, called *information-oriented signal-to-noise ratio (SNR)*

$$\mathrm{SNR}_b = \frac{E_b}{N_0} = \frac{E_c}{r \cdot 2\sigma^2} \tag{3.8}$$

where $E_b = E_c/r$ is the energy per *information* bit and $N_0 = 2\sigma^2$ is called the *double-sided power spectral density*. It can be shown that the $i$-th LLR value $\lambda_i$ of an AWGN channel is itself a normally distributed random variable,

$$\lambda_i \sim \mathcal{N}\left(4r(-1)^{x_i} \cdot \mathrm{SNR}_b, 8r \cdot \mathrm{SNR}_b\right), \tag{3.9}$$

hence the specific values of $E_b$ and $\sigma$ are irrelevant for the channel law.

In order to evaluate the performance of a specific code/decoder pair, it is common to state the frame-error rate not only for a single SNR, but instead to plot (3.4) for a whole range of SNR values. Since in the majority of cases the FER cannot be determined analytically, these *performance curves* are usually obtained by Monte Carlo simulation. To that end, (3.9) is utilized to generate a large number of channel outputs, until a sufficient number of decoding errors ($\textsc{decode}(y) \neq x$) allows for a statistically significant estimation of (3.4).

## 3.4 Binary Linear Block Codes

**3.4 Definition (linear code):** A binary $(n, k)$ code $\mathscr{C}$ is called *linear* if $\mathscr{C}$ is a linear subspace of $\mathbb{F}_2^n$. Consequently, a linear code admits a linear encoding function. ◁

Linear codes constitute the by far most important class of codes that are studied in literature. This is justified by the fact that, for binary-input symmetric memoryless channels, the results of Theorem 3.3 continue to hold when restricting to linear codes only [Gal68, Ch. 6.2].

Linearity implies a vast amount of structure and allows codes to be compactly defined by matrices, as introduced below. Note that all operations on binary vectors in this section are performed in $\mathbb{F}_2$, i.e., "modulo 2".

**3.5 Definition (dual code and parity-check matrices):** The orthogonal complement

$$\mathscr{C}^\perp = \left\{\xi \in \mathbb{F}_2^n \colon \xi^T x = 0 \text{ for all } x \in \mathscr{C}\right\}$$

of a linear $(n, k)$ code $\mathscr{C}$ is called the *dual code of* $\mathscr{C}$, the elements of $\mathscr{C}^\perp$ are *dual codewords* of $\mathscr{C}$.

A matrix $H \in \mathbb{F}_2^{m \times n}$ is a *parity-check matrix for* $\mathscr{C}$ if its rows generate $\mathscr{C}^\perp$ (i.e., the rows contain a basis of $\mathscr{C}^\perp$) and hence the equation

$$\mathscr{C} = \{x \colon Hx = 0\} \tag{3.10}$$

completely characterizes $\mathscr{C}$. In practice, $\mathscr{C}$ is often defined by stating a parity-check matrix $H$ in the first place; in that event, we also speak of $H$ as *the* parity-check matrix of $\mathscr{C}$. ◁

Since $\mathscr{C}^\perp$ is an $(n, n-k)$ code, it follows that $m > n - k$ in the above definition; in practice, $m = n - k$ is usually the case. Because $\mathscr{C}$ is linear, a linear *encoding* function can likewise be defined by means of a so-called *generator matrix* $G$, the rows of which form a basis of $\mathscr{C}$. Within the scope of this text, however, parity-check matrices are by far more important.

### 3.4.1 Minimum Hamming Distance

It is intuitively clear that, for a code to be robust against channel noise, the codewords should be maximally “distinguishable” from each other, i.e., there should be enough space between any two of them. Hence, one of the most important measures for the quality of a single code is its minimum distance, as defined below.

**3.6 Definition (minimum distance):** The *Hamming weight* $w_{\mathrm{H}}(x)$ of a binary vector $x$ is defined to be the number of 1s among $x$. The *Hamming distance* $d_{\mathrm{H}}(x,y) = w_{\mathrm{H}}(x-y)$ of two vectors $x, y$ of equal length is the number of positions in which they differ. The *minimum (Hamming) distance* of a linear code $\mathscr{C}$ is defined as

$$d_{\min}(\mathscr{C}) = \min_{\substack{x,y\in\mathscr{C}\\ x\neq y}} d_{\mathrm{H}}(x,y) = \min_{\substack{x,y\in\mathscr{C}\\ x\neq y}} |\{i\colon x_i \neq y_i\}|\,;$$

it is equivalent to the minimum Hamming weight among all non-zero codewords,

$$d_{\min}(\mathscr{C}) = \min_{x\in\mathscr{C}\setminus\{0\}} w_{\mathrm{H}}(x), \tag{3.11}$$

by linearity. ◁

The problem of finding the minimum distance of general linear codes is an NP-hard problem [Var97]. Nevertheless, integer programming techniques allow to compute $d_{\min}$ for codes which are not too large; see e.g. [Tan+10b; Sch+13]. In Section 4.4, with the *pseudoweight* we will encounter a similar weight measure that is specific to the LP decoding algorithm.

### 3.4.2 Factor Graphs

Let $\mathscr{C}$ be a binary linear code and $H \in \mathbb{F}_2^{m\times n}$ a parity-check matrix for $\mathscr{C}$. As noted before in (3.10), the condition

$$Hx = 0 \tag{3.12}$$

is necessary and sufficient for $x$ being a codeword of $\mathscr{C}$. A *row-wise* viewpoint of (3.12) leads to the following definition.

**3.7 Definition:** Let $\mathscr{C}$ be a linear $(n,k)$ code defined by the $m\times n$ parity-check matrix $H$. Any code $\mathscr{C}'$ such that $\mathscr{C} \subseteq \mathscr{C}'$ is called a *supercode* of $\mathscr{C}$. For $j \in \{1,\dots,m\}$, the particular supercode

$$\mathscr{C}_j = \{x\colon H_{j,\bullet}x = 0\}$$

is called the *$j$-th parity-check of $\mathscr{C}$*. It is a so-called *single parity check (SPC)* code, simply placing a parity condition on the entries $\{x_i\colon H_{j,i} = 1\}$. ◁

$$H = \begin{pmatrix} 1 & 1 & 0 & 1 & 1 & 0 & 0 \\ 0 & 1 & 1 & 1 & 0 & 1 & 0 \\ 0 & 0 & 0 & 1 & 1 & 1 & 1 \end{pmatrix}$$

(a) Parity-check matrix of the code.

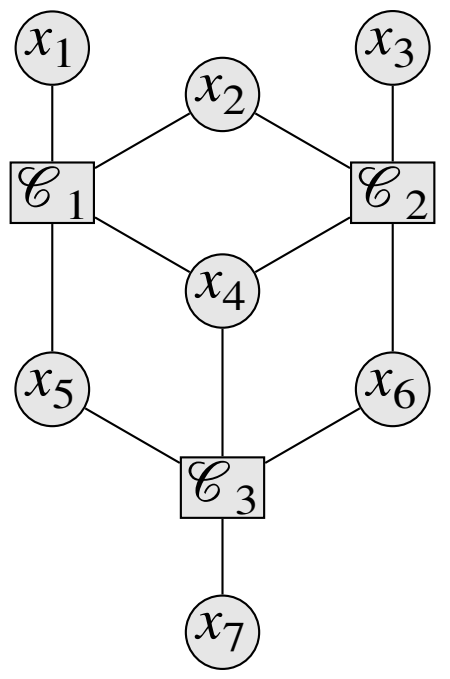

(b) Factor graph of the code.

Figure 3.5: Parity-check matrix and factor graph of a $(7, 4)$ code.

An obvious yet important consequence of the above definition is that

$$\mathscr{C} = \bigcap_j \mathscr{C}_j, \tag{3.13}$$

i.e., a linear code $\mathscr{C}$ is the intersection of the supercodes defined by the rows of a parity-check matrix for $\mathscr{C}$.

The fact that a linear code is characterized by several parity-check conditions placed on subsets of the variables is neatly visualized by a factor graph (or Tanner graph).

**3.8 Definition (factor graph):** The *factor graph* representing a parity-check matrix $H \subseteq \mathbb{F}_2^{m \times n}$ of a linear code $\mathscr{C}$ is a bipartite undirected graph $G = (V \dot\cup C, E)$ that has *m check nodes* $C = \{\mathscr{C}_1, \ldots, \mathscr{C}_m\}$, *n variable nodes* $V = \{x_1, \ldots, x_n\}$ and an edge $(\mathscr{C}_j, x_i)$ whenever $H_{j,i} = 1$. ◁

The factor graph representation plays an important role in the analysis and design of codes and decoding algorithms. One of today's most prominent decoding methods, named belief propagation, works by iteratively exchanging messages (representing momentary beliefs or probabilities) between the check and variable nodes, respectively, of the factor graph [KFL01]. Moreover, *graph covers* of the factor graph, as defined below in Section 4.4.3 have become an important tool to analyze LP decoding, belief propagation decoding, and their mutual relation [VK05].

## 3.5 Convolutional and Turbo Codes

*Turbo codes* constitute an important class of linear codes, as they were the first to closely approach the promises of Theorem 3.3 using a very efficient decoding algorithm [BG96].

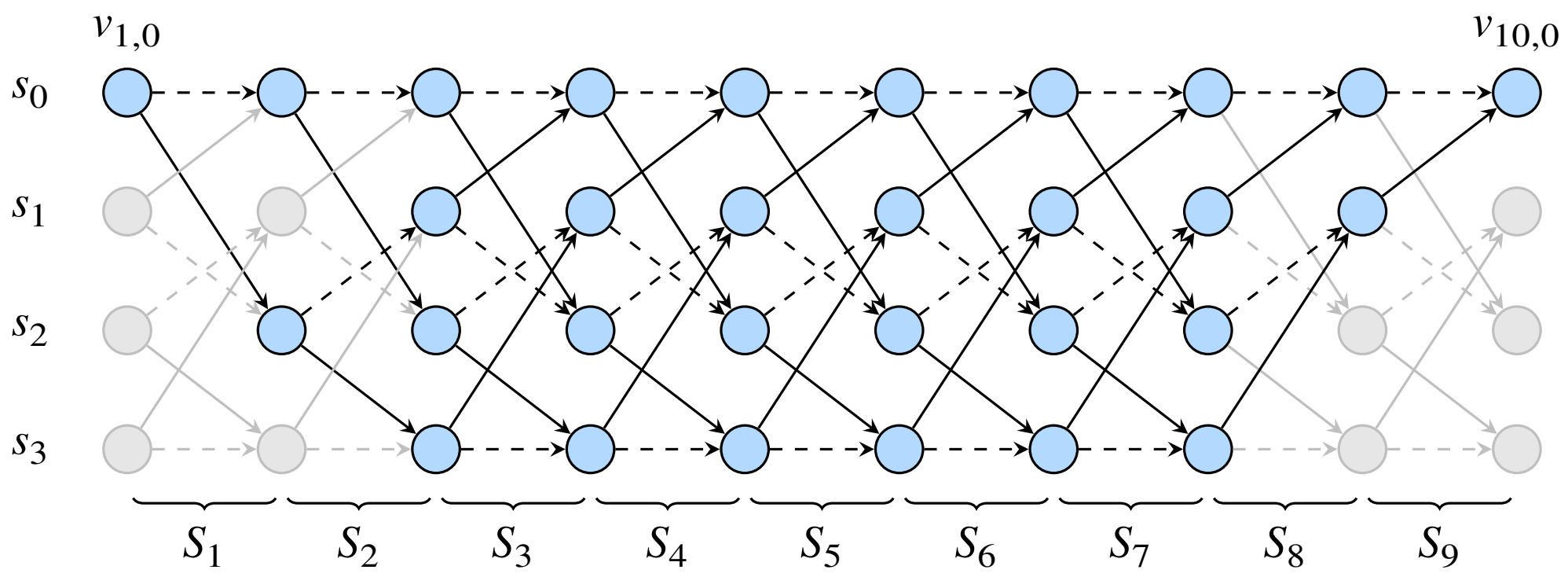

Figure 3.6: An example trellis graph with $k = 9$ segments and $2^d = 4$ states. Dashed edges have input bit $\text{in}(e) = 0$, for solid edges the input is $\text{in}(e) = 1$. Hence, for example, the zero input sequence $u = (0, \dots, 0)$ corresponds to the horizontal path $(v_{1,0}, v_{2,0}, \dots, v_{10,0})$ in $T$.

They are constructed by combining two or more (terminated) *convolutional codes*. For the matter of this work, it is important that the codewords of a convolutional code are in correspondence to certain *paths in a graph*, as described below.

A terminated convolutional $(n, k)$ code $\mathscr{C}$ with rate $r$ (where $1/r = n/k \in \mathbb{N}$) and *memory* $d \in \mathbb{N}$ can be compactly described by a finite state machine (FSM) with $2^d$ states $S = \{s_0, \dots, s_{2^d-1}\}$ and a state transition function $\delta\colon S \times \mathbb{F}_2 \to S \times \mathbb{F}_2^{1/r}$ that defines the encoding of an information word $u \in \mathbb{F}_2^k$ as follows. An example FSM is shown in Figure 3.7.

Initially, the FSM is in state $s^{(1)} = s_0$. Then the bits $u_i$ of $u$ are subsequently fed into the FSM to determine the codeword $\mathscr{C}(u)$, i.e., in each step $i \in \mathbb{N}$, the current state $s^{(i)} \in S$ together with the $i$-th input bit $u_i$ determines via

$$\delta(s^{(i)}, u_i) = (s^{(i+1)}, x^{(i)})$$

the subsequent state $s^{(i+1)}$ as well as $n/k$ output bits $x^{(i)} = (x_1^{(i)}, \dots, x_{n/k}^{(i)})$ that constitute the part of the codeword that belongs to $u_i$. Finally, the machine has to terminate in the zero state, i.e., $s^{(k+1)} = s_0$ is required (this entails that some of the input bits are not free to choose and thus have to be considered as part of the *output* instead; in favor of a clear presentation, however, we ignore this inexactness and assume that $u$ is in advance chosen such that $s^{(k+1)} = s_0$; see e.g. [HR13b] for a more rigorous construction). The encoded word $x = \mathscr{C}(u)$ now consists of a concatenation of the $x^{(i)}$, namely,

$$x = \left(x_1^{(1)}, \dots, x_{n/k}^{(1)}, \dots, x_1^{(k)}, \dots, x_{n/k}^{(k)}\right).$$

The FSM of a convolutional code is always defined in such a way that this encoding is a *linear* map, and hence $\mathscr{C}$ is a linear code.

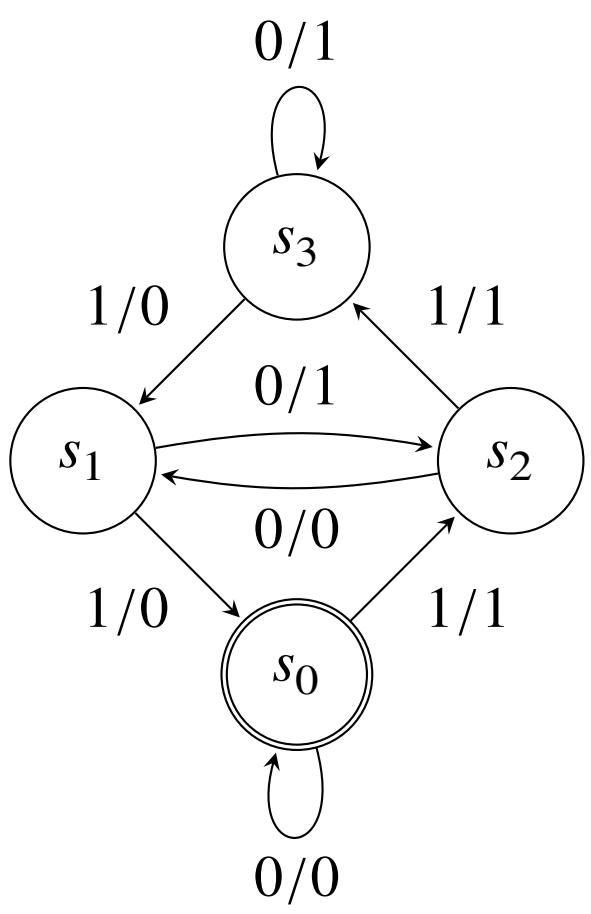

Figure 3.7: Finite state machine of a rate-1 convolutional code. The edge labels $u/x$ specify the respective input ($u$) and output ($x$) bits.

By "unfolding" the FSM along the time domain, we now associate a directed acyclic graph $T = (V, E)$ to the convolutional code $\mathscr{C}$, called its *trellis* (see Figure 3.6). Each vertex of $T$ corresponds to a state of the FSM at a specific time step, such that $V = \{1, \dots, k + 1\} \times S$, where we denote the vertex $(i, s)$ corresponding to state $s \in S$ at step $i$ shortly by $v_{i,s} \in V$.

The edges of $T$ in turn correspond to valid state transitions. For each $i \in \{1, \dots, k\}$ and $s \in S$, there are two edges emerging from $v_{i,s}$, according to the two possible values of $u_i$ (which are encoded in the *input labels* $\mathrm{in}(e) \in \{0, 1\}$ of the edges); both their *output labels* $\mathrm{out}(e) \in \mathbb{F}_2^{n/k}$ and target vertices $v_{i+1,s'}$ are determined by the state transition function via

$$\delta(s, \mathrm{in}(e)) = (s', \mathrm{out}(e)).$$

Hence, each edge $e = (v_{i,s}, v_{i+1,s'})$ of $T$ corresponds to the input of one bit at a specific step $i$ and a specific state $s$ of the encoder FSM that is in state $s'$ afterwards; the labels of $e$ define the value of the input bit $u_i = \mathrm{in}(e)$ and the output sequence $x^{(i)} = \mathrm{out}(e)$, respectively. The trellis $T$ is thus "$(k + 1)$-partite" in the sense that $V$ partitions into $k + 1$ subsets $V_i$ such that edges only exist between two subsequent sets $V_i$ and $V_{i+1}$. This motivates the definition of the $i$-th trellis *segment* $S_i = (V_i \cup V_{i+1}, E_i)$ according to the $i$-th encoding step as the subgraph induced by $V_i \cup V_{i+1}$.

The transition function $\delta$ is always designed in such a way that if $\delta(s, 0) = \delta(s', x')$ and $\delta(s, 1) = \delta(s'', x'')$ then $x' \neq x''$, i.e., at each encoding step, the two outputs corresponding to an input bit 0 and 1, respectively, must be different. As a consequence, the codewords of $\mathscr{C}$ are in one-to-one correspondence with the paths from $v_{1,0}$ to $v_{k+1,0}$ in $T$: at step $i$ in state $s$, the next $n/k$ bits of the codeword determine which edge to follow from $v_{i,s}$, while conversely the output label of such an edge fixes the next $n/k$ code bits. Due to the boundary constraints $s^{(1)} = s^{(k+1)} = s_0$, some vertices and edges in the leading as well as the endmost $d$ segments are not part of any such path and therefore are usually removed from $T$, as shown in the figure.

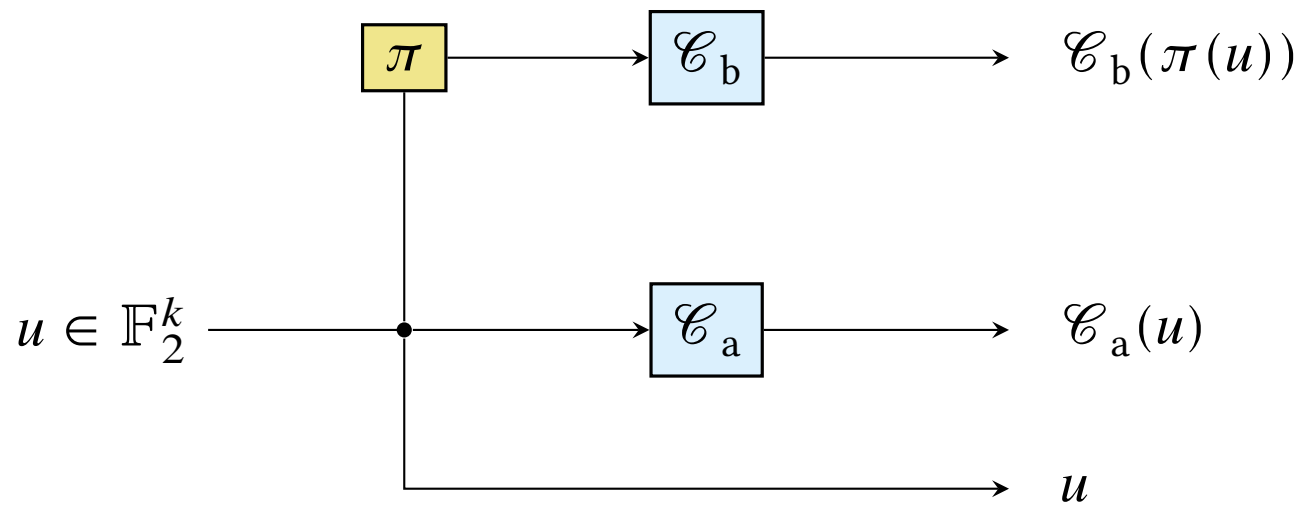

Figure 3.8: Encoding scheme of a turbo code, consisting of two parallely concatenated convolutional codes $\mathscr{C}_\mathrm{a}$ and $\mathscr{C}_\mathrm{b}$, respectively.

In a turbo code $\mathscr{C}_\mathrm{TC}$, now, several convolutional codes are *concatenated* in order to improve upon the rather weak error-correction performance of plain convolutional codes. In the most common form, two identical convolutional codes $\mathscr{C}_\mathrm{a}$ and $\mathscr{C}_\mathrm{b}$ with rate $r = 1$ each are concatenated *parallely*, which means that the information word $u$ is encoded by both, but the entries of $u$ are permuted by a fixed permutation (the *interleaver*) $\pi \in \mathbb{S}_k$ before entering the second component code $\mathscr{C}_\mathrm{b}$ (see Figure 3.8). A codeword of the turbo code $\mathscr{C}_\mathrm{TC}$ then consists of the concatenation of a copy of $u$, $\mathscr{C}_\mathrm{a}(u)$ and $\mathscr{C}_\mathrm{b}(\pi(u))$, so that the overall rate of $\mathscr{C}_\mathrm{TC}$ is $r = 1/3$ (here, again, a small rate loss due to termination is ignored). In a more general setting, the term *turbo-like codes* refers to schemes that include *serial* concatenation, where the output of one convolutional code is used as input for another convolutional code, or any combination of parallel and serial concatenations.

Taking the path representation of codewords of $\mathscr{C}_\mathrm{a}$ and $\mathscr{C}_\mathrm{b}$ in their respective trellis graphs $T_\mathrm{a}$ and $T_\mathrm{b}$ into account, from the above definition of a turbo code $\mathscr{C}_\mathrm{TC}$ we can derive a one-to-one correspondence between codewords of $\mathscr{C}_\mathrm{TC}$ and pairs $(P_\mathrm{a} = (e_1^\mathrm{a}, \dots, e_k^\mathrm{a}), P_\mathrm{b} = (e_1^\mathrm{b}, \dots, e_k^\mathrm{b}))$ of paths in $T_\mathrm{a}$ and $T_\mathrm{b}$, respectively, which additionally fulfill that

$$\mathrm{in}(e_i^\mathrm{a}) = \mathrm{in}(e_{\pi(i)}^\mathrm{b}), \tag{3.14}$$

i.e., the $i$-th edge in $P_\mathrm{a}$ must have the same input label as the $\pi(i)$-th edge in $P_\mathrm{b}$, because both equal the $i$-th input bit $u_i$. The application of this path–code relationship to decoding by mathematical optimization is introduced in Section 4.5.

## 3.6 Non-Binary Codes and Higher-Order Modulation

So far, we assumed *binary* data processing throughout coding and data transmission, as introduced in Section 3.1. While today's microelectronic systems, as is generally known,

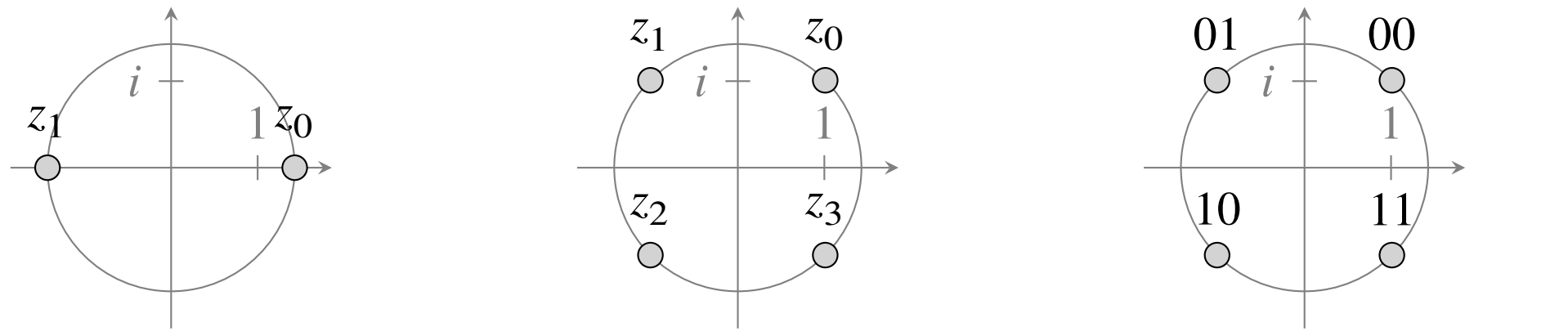

(a) Binary modulation.

(b) Modulation scheme using $Q = 4$ different complex signals.

(c) Example mapping of $k = 2$ bits to $4 = 2^k$ complex symbols.

Figure 3.9: Binary and non-binary modulation schemes.

*internally* rely on the binary representation of data, the above simplification is, in two different but related ways, not the whole story in case of channel coding.

First, observe that the definition of (linear) codes can straightforwardly be generalized to any finite field $\mathbb{F}_q$ for a prime power $q$: the information and codewords still lie in vector spaces and linear maps can be defined as usual; the parity-check matrix of such a *non-binary code* then has entries in $\{0, \dots, q-1\}$. Several constructions of strong codes rely on a non-binary field, thus a restriction to the binary case would prevent us from using those codes.

Secondly, in many practical transmission systems the signal space is modeled by the *complex* plane, where the real and imaginary axis, respectively, correspond to two different carrier waves (e.g., two sines that are out of phase by $\pi/2$) such that any complex number $z$ represents a linear combination of both waves that is then emitted onto the carrier medium. This technique is called *modulation*. At the receiver's side, a complementary *demodulator* measures the potentially distorted wave and reconstructs (e.g. via Fourier analysis) a point $\tilde{z}$ in the complex plane.

In the most simple *binary* case, two complex numbers (e.g. $z_0 = 1 + 0i$ and $z_1 = -1 + 0i$) are chosen that represent the values 0 and 1, respectively, of a single bit. If we now assume that the channel adds independent Gaussian noise to both carrier waves, this case reduces to the binary AWGN channel as introduced in Section 3.3.

In *higher-order modulation*, however, more than one bit of information is transmitted at once by choosing $Q > 2$ (usually, $Q$ is a power of 2) possible complex signals $\{z_0, \dots, z_{Q-1}\}$. Hence, the channel can be modeled by $Q$ probability functions $P(\tilde{z} \mid 0), \dots, P(\tilde{z} \mid Q-1)$ where $\tilde{z} \in \mathbb{C}$. Figure 3.9 shows binary and one example non-binary modulation scheme in the complex plane.

Non-binary codes and higher-order modulation can be combined in several ways. Most obviously, if $q = Q$ then to each complex symbol $z_0, \dots, z_{Q-1}$ an element of $\mathbb{F}_q$ can be assigned, such that the channel transmits one entry of the codeword per signal. On the other hand, if e.g. $q = 2$ and $Q = 2^k$, $k$ bits of the codeword can be sent *at once* by mapping the $2^k$ possible bit configurations to the $2^k$ chosen complex symbols.

In both cases, the expression for ML decoding (3.6) is more complex than in the binary case. In particular, the linearization to formulate ML decoding as an IP (Section 4.1) is not possible in the same way because there is no individual LLR value corresponding to each channel signal. Nevertheless, ML decoding of non-binary codes was formulated as an IP in [Fla+09], and the incorporation of higher-order modulation into the IP model is covered in [Sch+12].

# 4 Decoding by Optimization: The Connection

So far, the two subjects introduced above—linear and integer optimization in Chapter 2 on the one hand and coding theory in Chapter 3 on the other—may seem to have little in common: while the first consists of a mixture of (linear) algebra and probability, the latter is concerned with solution algorithms for specific linear or discrete problems. The two areas become linked, however, by the observation that *de*coding a received signal, in particular the (optimal) ML decoding as introduced in Section 3.2, amounts to solving a combinatorial optimization problem that can be formulated as an IP.

Therefore, this section introduces the abovementioned connection by reviewing the IP formulation of ML decoding and is then mainly concerned with a particular LP relaxation of that formulation, called *LP decoding*. The style of writing is intentionally a little more verbose than it was in the two previous chapters because, first, this part is the most probable for the audience to be unfamiliar with and, secondly, due to the recency of the subject, we are not aware of any up-to-date, tutorial-like, yet mathematically stringent document that covers what we believe to be its most important aspects.

A well-written resource for LP decoding is the dissertation of its inventor Feldman [Fel03]. Large parts of Section 4.4 are elaborately presented in [VK05] which is abounding in examples. Not least, [HRT12] includes a literature survey of the algorithmic aspects of optimization-based decoding until the time of its writing, as well as a short coverage of the underlying theory.

## 4.1 ML Decoding as Integer Program

While the perception of ML decoding (at least on the BSC) as a combinatorial optimization problem is probably as old as coding theory itself (for example, the proof of its NP-hardness by Berlekamp, McEliece, and Tilborg in 1978 [BMT78] constitutes an obvious connection to the optimization community), it has been only in 1998 that an *integer programming* formulation of the problem was given [Bre+98], which consists of linearizations (with respect to $\mathbb{R}$) of both the objective function and the code structure. In the following, we present a slightly modified version (as in [Tan+10a]) of that construction.

Recall that the ML codeword maximizes the likelihood function $P(y \mid x)$ for a received channel output $y$ (3.6). Since the channel is assumed to be memoryless, we have [Bre+98; Fel03]

$$\hat{x}_{\text{ML}} = \arg\max_{x \in \mathscr{C}} \prod_{i=1}^{n} P(y_i \mid x_i) \tag{4.1a}$$

$$= \arg\min_{x \in \mathscr{C}} -\sum_{i=1}^{n} \ln P(y_i \mid x_i) \tag{4.1b}$$

$$= \arg\min_{x \in \mathscr{C}} \sum_{i=1}^{n} (\ln P(y_i \mid 0) - \ln P(y_i \mid x_i)) \tag{4.1c}$$

$$= \arg\min_{x \in \mathscr{C}} \sum_{i:\, x_i = 1} \ln \left( \frac{P(y_i \mid 0)}{P(y_i \mid 1)} \right) \tag{4.1d}$$

Since the fraction in the last term exactly matches the LLR value $\lambda_i$ (3.2) which is known to the observer, we see that ML decoding is equivalent to minimizing the linear functional $\lambda^T x$ over all codewords $x \in \mathscr{C}$.

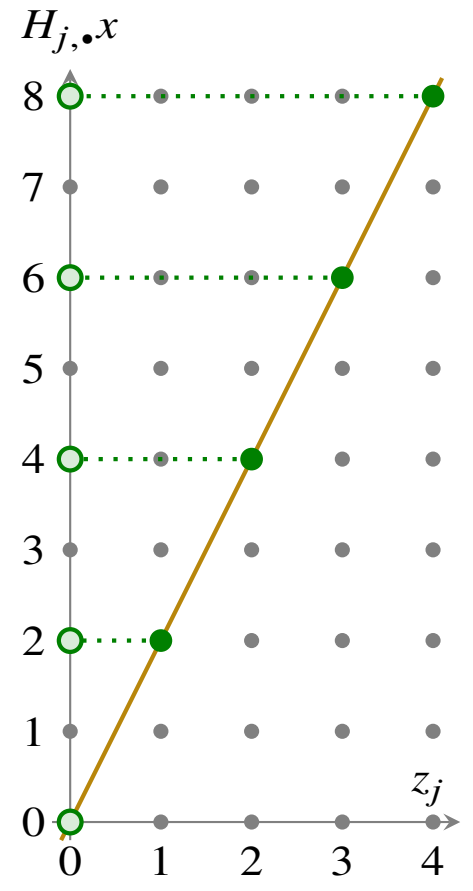

Figure 4.1: Feasible integer points of the linearization $H_{j,\bullet}x - 2z_j = 0$.

How can we grasp this condition "$x \in \mathscr{C}$" by an IP? The answer lies in the code-defining equation $Hx = 0$ (3.12) for a given parity-check matrix $H \in \mathbb{F}_2^{m \times n}$, which can be $\mathbb{R}$-linearized in virtue of *auxiliary* integer variables $z \in \mathbb{Z}^m$ as follows: The condition $x \in \mathscr{C}$ is eqivalent to $Hx = 0 \pmod 2$, which in turn is fulfilled if and only if the result of $Hx$, as an operation in the reals, is a vector whose entries are even numbers. It is thus clear that the formulation

$$\min \quad \lambda^T x \tag{4.2a}$$

$$\text{s.t.} \quad Hx - 2z = 0 \tag{4.2b}$$

$$x \in \mathbb{F}_2^n,\ z \in \mathbb{Z}^m \tag{4.2c}$$

models ML decoding because (4.2b) can be achieved by an *integral* vector $z$ if and only if $Hx$ is even (see Figure 4.1).

Note that any IP formulation of the ML decoding problem can be easily modified to output the minimum distance $d_{\text{min}}$ of a code: in view of (3.11), this is equivalent to determine a codeword of minimum Hamming weight. By setting $\lambda = (1, \dots, 1)$, the objective function value (4.2a) equals the Hamming weigth of $x$, and an additional linear constraint $\sum x_i \geq 1$ excludes the all-zero codeword, such that the IP solution must be a codeword of minimum Hamming weight ([Pun+10; KD10]).

Interestingly, the above linearization of the code was apparently "forgotten" and several years later reinvented in 2009 [Tan+09] and 2010 [KD10]. One possible explanation might be that while (4.2) is very compact in terms of size, its LP relaxation is essentially useless: if (4.2c) is replaced by its continuous counterpart, then the feasible region of the $x$ variables is the entire unit hypercube—for any configuration of $x$, a corresponding real $z$ can be found

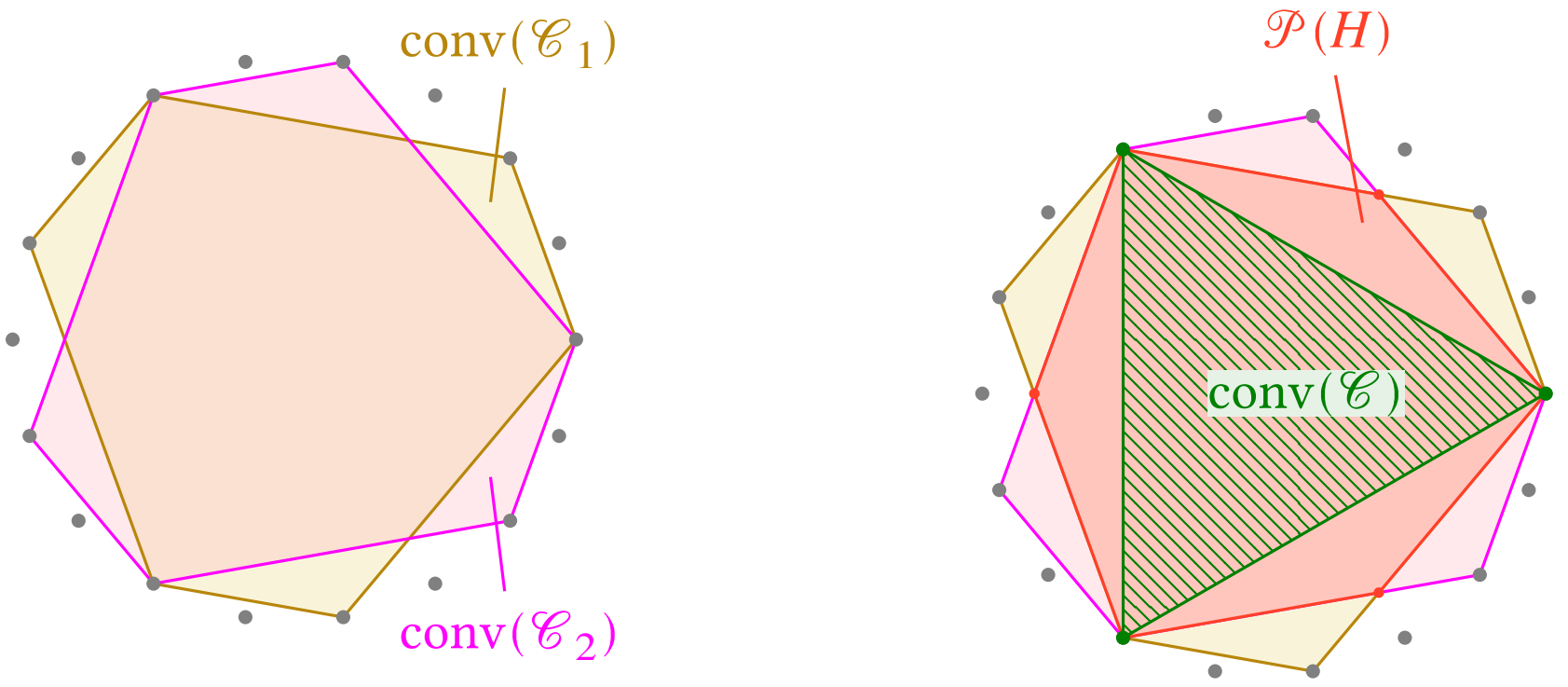

(a) Two constitutent polytopes $\mathrm{conv}(\mathscr{C}_1)$ and $\mathrm{conv}(\mathscr{C}_2)$.

(b) $\mathscr{P}(H) = \mathrm{conv}(\mathscr{C}_1) \cap \mathrm{conv}(\mathscr{C}_2)$ is a superset of $\mathrm{conv}(\mathscr{C})$ with additional fractional pseudocodewords $\notin \mathbb{F}_2^n$.

Figure 4.2: Sketch of $\mathscr{P}(H) = \bigcap_j \mathrm{conv}(\mathscr{C}_j)$ as a true subset of $\mathrm{conv}(\mathscr{C}) = \mathrm{conv}\left(\bigcap_j \mathscr{C}_j\right)$. In this geometrically incorrect sketch, the circular dots represent $\mathbb{F}_2^n$.

such that (4.2b) is fulfilled. It should be noted, however, that the formulation (4.2) has found recent justification by the fact that it appears to perform very well with commercial IP solvers [Tan+10b; Pun+10].

## 4.2 LP Decoding

It was the *LP decoder* introduced by Feldman [Fel03; FWK05] that established linear programming in the field of decoding by providing several equivalent IP formulations for which even the LP relaxations exhibit a decoding performance that is of interest for practical considerations.

The essence of Feldman's LP decoder lies in the representation (3.13) of a code, together with the fact that for the SPC codes $\mathscr{C}_j$ (Definition 3.7), a polynomially sized description of $\mathrm{conv}(\mathscr{C}_j)$ by means of (in)equalities and potential auxiliary variables is possible. Instead of providing an LP description of $\mathrm{conv}(\mathscr{C})$ (which in view of the NP-hardness of ML decoding is unlikely to be tractable), the LP decoder thus operates on the relaxation polytope

$$\mathscr{P}(H) = \bigcap_j \mathrm{conv}(\mathscr{C}_j) \supseteq \mathrm{conv}(\mathscr{C}), \tag{4.3}$$

called the *fundamental polytope* [VK05] of the parity-check matrix $H$. The vertices of $\mathscr{P}(H)$ are also called *pseudocodewords*. Note that the "$\supseteq$" in the above expression is usually strict; the sketch in Figure 4.2 might help to realize why this is the case.

**4.1 Definition (LP decoder):** Let $H$ be a parity-check matrix for the linear code $\mathscr{C}$ and $\lambda$ the vector of channel LLR values. The *LP decoder* LP-DECODE($\lambda$) outputs, for given $\lambda \in \mathbb{R}^n$, the optimal solution $\hat{x}$ of the LP

$$\min \quad \lambda^T x \tag{4.4a}$$
$$\text{s.t.} \quad x \in \mathscr{P}(H), \tag{4.4b}$$

where $\mathscr{P}(H)$ is as defined in (4.3). ◁

The above definition is a meaningful relaxation of conv($\mathscr{C}$) because one can easily show that $\mathscr{P}(H) \cap \{0,1\}^n = \mathscr{C}$, i.e., the codewords of $\mathscr{C}$ and the integral vertices of $\mathscr{P}(H)$ coincide, which proves the following theorem.

**4.2 Theorem (ML certificate [Fel03]):** *The LP decoder has the* ML certificate property*:*

$$\hat{x} = \text{LP-DECODE}(\lambda) \in \{0,1\}^n \Rightarrow \hat{x} = x_{\text{ML}},$$

*i.e., if $\hat{x}$ is integral, it must be the ML codeword. Put another way, solving (4.4) as an* IP *with the additional constraint $x \in \{0,1\}^n$ constitutes a true ML decoder.* ◁

Note that if we had $\mathscr{P}(H) = \text{conv}(\mathscr{C})$, the LP decoder would actually be an ML decoder. Because this is not the case in general (moreover, it apparently does not hold for any interesting code; see [Kas08]), the inclusion $\text{conv}(\mathscr{C}) \subsetneq \mathscr{P}(H)$ is usually strict, and the difference $\mathscr{P}(H) \setminus \text{conv}(\mathscr{C})$ must be due to additional *fractional* vertices of $\mathscr{P}(H)$, i.e., vectors for which at least one entry is neither 0 nor 1.

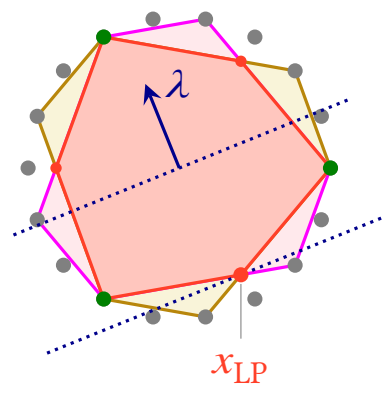

Figure 4.3: LLR vector $\lambda$ for which $x_{\text{LP}} \notin \mathbb{F}_2^n$ is optimal .

Feldman gave three different formulations of conv($\mathscr{C}_j$), the convex hull of the SPC codes constituting $\mathscr{C}$, to be used in (4.4b). In the context of this work, only the one described below, which is named $\Omega$ in [Fel03] and based on [Jer75], is relevant.

**4.3 Theorem:** *Let $H$ and $\mathscr{C}$ as above and let $N_j = \{i : H_{j,i} = 1\}$ be the indices covered by the $j$-th parity check $\mathscr{C}_j$ of $\mathscr{C}$. Then the inequalities*

$$\sum_{i \in S} x_i - \sum_{i \in N_j \setminus S} x_i \le |S| - 1 \qquad \textit{for all } S \subseteq \mathscr{N}_j \textit{ with } |S| \textit{ odd} \tag{4.5a}$$
$$0 \le x_i \le 1 \qquad \textit{for } i = 1, \ldots, n \tag{4.5b}$$

*precisely define the convex hull of $\mathscr{C}_j$.* ◁

As each inequality (4.5a) explicitly forbids one odd-sized set $S$, i.e., a configuration for which $H_{j,\bullet} x \equiv 1 \pmod 2$ (it is violated by a binary vector $x$ if and only if $x_i = 1$ for $i \in S$ and $x_i = 0$ for $i \in N_j \setminus S$), they are also called *forbidden-set inequalities.* Note that the number of such inequalities is exponential in the size of $N_j$, which is why LP decoding was first proposed for codes defined by a *sparse* matrix $H$, so-called *LDPC* codes [Gal60; Mac99]. It will however turn out in the following review of adaptive LP decoding that the inequalities (4.5a) can be efficiently *separated*, which renders their exponential quantity harmless in practice.

## 4.3 Adaptive LP Decoding

The prohibitive size of the LP decoding formulation (4.4), especially for dense $H$ and larger block lengths, can be overcome by a cutting plane algorithm (cf. Section 2.3.1), called *adaptive LP decoding*, as proposed in [TS08; TSS11]. It starts with the trivial problem of minimizing the objective function over the unit hypercube:

$$\begin{aligned} \min \quad & \lambda^T x \\ \text{s.t.} \quad & x \in [0,1]^n \end{aligned}$$

and then iteratively *refines* the domain of optimization by inserting those forbidden-set inequalities (4.5a) that are *violated* by the current solution, and hence constitute valid *cuts*; see Figures 4.4(a) to 4.4(c) for a sketch. The procedure to find a cut in the $j$-th row of $H$ (it is shown in [TS08] that, at any time, one row of $H$ can provide at most one cut) is based on the following reformulation of (4.5a):

$$\sum_{i \in S} (1 - x_i) + \sum_{i \in N_j \setminus S} x_i \geq 1. \tag{4.6}$$

To find a violating inequality (if it exists) of the form (4.6), an odd-sized set $S$ needs to be found that minimizes the left-hand side of (4.6). It is easy to show [TSS11] that this can be accomplished by taking all $i$ with $x_i > 1/2$ and, if that set is even-sized, remove or add the index $i^*$ for which $x_{i^*}$ is closest to $1/2$.

When no more violating inequalities are found, the solution equals that of (4.4) and the algorithm terminates. The total number of inequalities in the final model is however bounded by $n^2$, which shows that the adaptive approach indeed overcomes, with respect to size, the problems of the model (4.5).

An important advantage of the separation approach is that one can immediately incorporate *additional* types of cutting planes—if it is known how to solve the corresponding separation problem, i.e., find violated cuts from the current LP solution—in order to tighten the LP relaxation (4.4). A successful method of doing so is by using redundant parity-checks.

**4.4 Definition:** Let $\mathscr{C}$ be a linear code defined by a parity-check matrix $H$. A dual codeword $\xi \in \mathscr{C}^\perp$ that does not appear as a row of $H$ is called a *redundant parity-check (RPC)*. An RPC $\xi$ is said to *induce a cut* at the current LP solution $x$ if one of the inequalities (4.5a) derived from $\xi$ is violated by $x$. ◁

RPCs are called "redundant" because the rows of $H$ already contain a basis of $\mathscr{C}^\perp$ by definition, thus every RPC must be the (modulo-2) sum of two or more rows of $H$. The following result [Tan+10a] gives a strong clue which RPCs might potentially induce cuts.

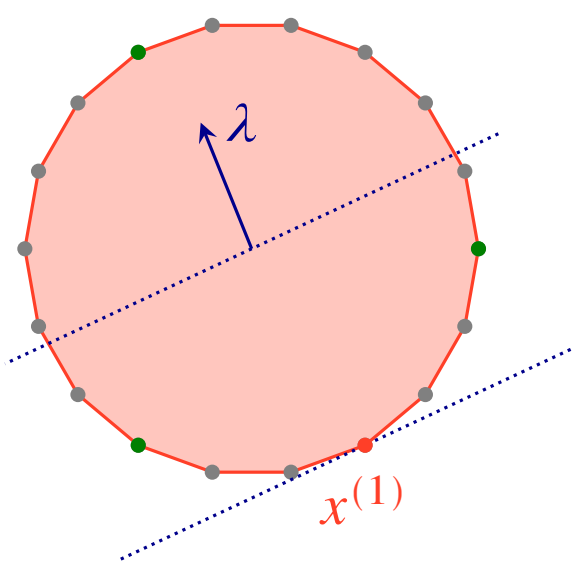

(a) Initial optimization over $[0, 1]^n$.

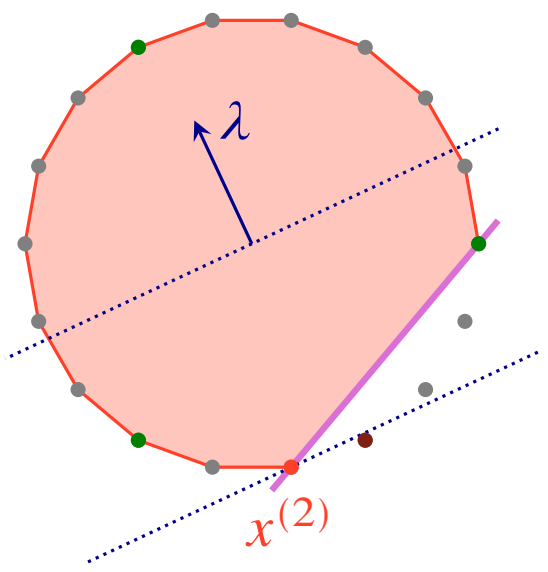

(b) cut for $x^{(1)}$ leads to next solution $x^{(2)}$.

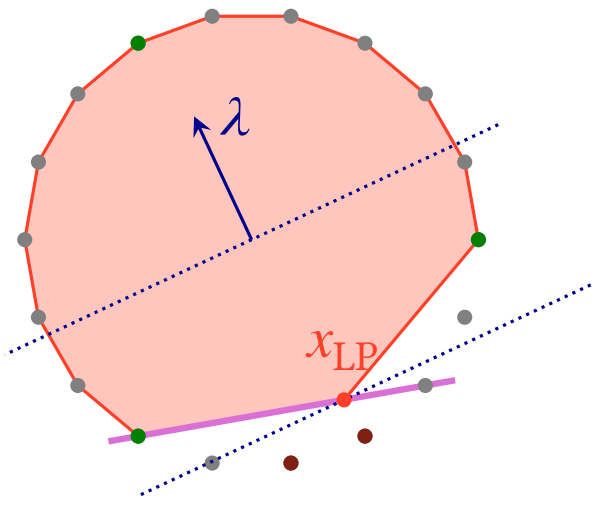

(c) Another cut yields the same solution $x_{\mathrm{LP}}$ as before.

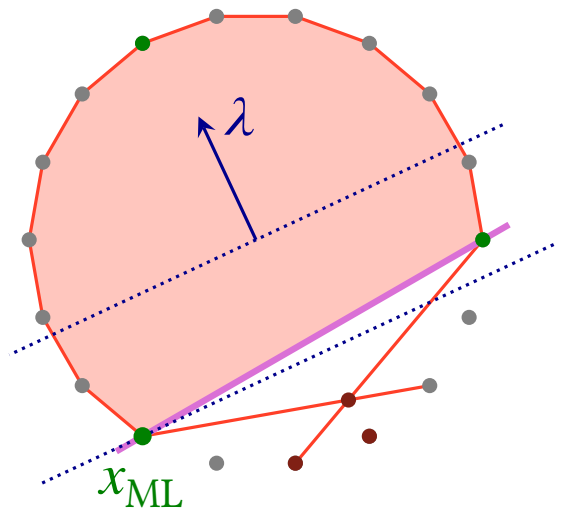

(d) An example RPC cut (in the picture, the cut is a facet of $\mathrm{conv}(\mathscr{C})$) that would lead from the above LP solution to the ML codeword $x_{\mathrm{ML}}$.

Figure 4.4: Sketch of the execution of adaptive LP decoding, based on the instance shown in Figure 4.3.

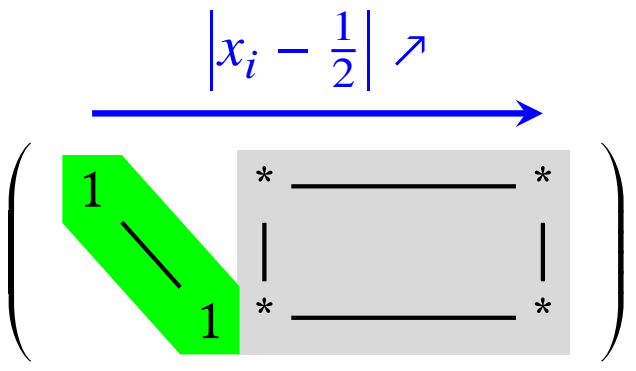

Figure 4.5: Structure of the alternative parity-check matrix $\bar{H}$ obtained from RPC cut search: diagonalized part at the left after reordering of columns by $\left|x_i - \frac{1}{2}\right|$.

**4.5 Lemma:** *Let $\xi \in \mathscr{C}^\perp$ be a dual codeword and $x$ an intermediate solution of the adaptive LP decoding algorithm. If*

$$|\{i\colon \xi_i = 1 \text{ and } x_i \notin \{0,1\}\}| = 1,$$

*i.e., exactly one index of the fractional part of $x$ is contained in the support of $\xi$, then $\xi$ induces an RPC cut for $x$.* ◁

An efficient method to search for RPC cuts in view of the above observation works as follows [Tan+10a; ZS12] (see Figure 4.5): Given an intermediate LP solution $x$,

(1) sort the columns of $H$ according to an ascending ordering of $|x_i - 1/2|$,

(2) perform Gaussian elimination on the reordered $H$ to diagonalize its leftmost part, resulting in an alternative parity-check matrix $\tilde{H}$, then

(3) search for cuts among the rows of $\tilde{H}$ as in adaptive LP decoding.

The motivation behind this approach is that, if the submatrix of $H$ corresponding to the fractional part of $x$ has full column rank, the leftmost part of $\tilde{H}$ will be a diagonal matrix, and hence by Lemma 4.5 *every row* of $\tilde{H}$ would induce a cut for $x$. The results reported in [Tan+10a] and [ZS12] furthermore suggest that, even if this is not the case and thus the requirements of Lemma 4.5 are not necessarily met, this "sort-&-diagonalize" strategy very often leads to cuts and substantially improves the error-correcting capability of the plain LP decoder that does not involve RPCs (see Figure 4.4(d)).

## 4.4 Analysis of LP Decoding

While the aspects of LP decoding discussed so far include some useful theoretical results about an *individual* run of the algorithm (most importantly, the ML certificate property given in Theorem 4.2), there is no immediate theoretical approach to determine the *average* error-correction performance (3.4) of a given code and channel under LP decoding, other than using simulations as described in Section 3.3.

In the following, we briefly outline an approach to a theoretical performance analysis of LP decoding that is based on a channel-specific rating of the vertices of the LP decoding polytope, called *pseudoweight*. The theory presented in this section is based on the “plain” LP decoder as defined in Definition 4.1, i.e., does not take the improvement via RPC cuts (Definition 4.4 and the discussion thereafter) into account; most of the results can however be extended to that case in a straightforward manner.

### 4.4.1 All-Zero Decoding and the Pseudoweight

First we introduce the very useful *all-zeros assumption*.

**4.6 Theorem (Feldman [Fel03]):** *If the LP decoder (4.4) is used on a binary-input memoryless symmetric channel, the probability of decoding error is independent of the sent codeword: the FER (3.4) satisfies*

$$\text{FER} = P(\text{LP-}\textsc{decode}(\lambda) \neq 0 \mid 0 \textit{ was sent}). \qquad \triangleleft$$

The proof of Theorem 4.6 relies on the symmetry of both the channel and the polytope. The latter is due to the linearity of the code (which implies that $\text{conv}(\mathscr{C})$ basically “looks the same” from any codeword $x$) and the $\mathscr{C}$-symmetry [Fel03, Ch. 4.4] of $\mathscr{P}(H)$, which extends that symmetry to the relaxed LP polytope. As a consequence of the theorem, when examining the LP decoder’s error probability we can always assume that the all-zero codeword $0 \in \mathscr{C}$ was sent, which greatly simplifies analysis.

Assume now that the all-zero codeword is sent through a channel and the result $\lambda$ is decoded by the LP decoder which solves (4.4) to obtain the optimal solution $\hat{x}$. The decoder fails if there is a vertex $x$ of $\mathscr{P}(H)$ such that

$$\lambda^T x < \lambda^T 0 = 0 \tag{4.7}$$

(we assume here and in the following that in case of ties, i.e., $\lambda^T x = 0$ for some non-zero vertex $x$, the LP decoder correctly outputs 0; for the AWGN channel, ties can be neglected since they occur with probability 0). The probability $P(\lambda^T x < 0)$ of the event (4.7), also called the *pairwise error probability* between $x$ and 0, depends on the channel. In case of the AWGN channel, by (3.9) we have

$$\lambda_i x_i \sim \mathcal{N}\left(4r \cdot \text{SNR}_\text{b} x_i, 8r \cdot \text{SNR}_\text{b} x_i^2\right)$$

and because the channel treats symbols independently and furthermore the sum of independent Gaussian variables is again gaussian with mean and variance simply summing up, we obtain

$$\lambda^T x \sim \mathcal{N}\left(4r \cdot \text{SNR}_\text{b} \|x\|_1, 8r \cdot \text{SNR}_\text{b} \|x\|_2^2\right). \tag{4.8}$$

Hence, $\lambda^T x$ is again Gaussian, and the probability that $\lambda^T x < 0$ computes as (using the abbreviations $\mu = 4r \cdot \mathrm{SNR_b} \, \|x\|_1$ and $\sigma^2 = 8r \cdot \mathrm{SNR_b} \, \|x\|_2^2$)

$$P(\lambda^T x < 0) = \frac{1}{\sqrt{2\pi\sigma^2}} \int_{-\infty}^{0} e^{-\frac{(x-\mu)^2}{2\sigma^2}} \mathrm{d}x = \frac{1}{\sqrt{2\pi}} \int_{\frac{\mu}{\sigma}}^{\infty} e^{-\frac{1}{2}x^2} \mathrm{d}x.$$

Introducing the $Q$-function as $Q(a) = \int_a^\infty \frac{1}{\sqrt{2\pi}} e^{-\frac{x^2}{2}} \mathrm{d}x$, we get

$$P(\lambda^T x < 0) = Q\left(\frac{\mu}{\sigma}\right) = Q\left(\sqrt{2r \cdot \mathrm{SNR_b} \frac{\|x\|_1^2}{\|x\|_2^2}}\right) = F\left(\frac{\|x\|_1^2}{\|x\|_2^2}\right),$$

for a monotone function $F$, which motivates the following definition.

**4.7 Definition ([For+01; VK05]):** Let $x$ be a non-zero vertex of $\mathscr{P}(H)$. The *(AWGN) pseudoweight* of $x$ is defined as

$$w_{\mathrm{p}}^{\mathrm{AWGN}}(x) = \|x\|_1^2 \big/ \|x\|_2^2 \,. \tag{4.9}$$

◁

Observe that the pairwise error probability $P(\lambda^T x < 0)$ is a strictly monotonically decreasing function of $w_{\mathrm{p}}^{\mathrm{AWGN}}(x)$: the lower the pseudoweight is, the higher is the probability that the LP decoder wrongly runs into $x$ instead of 0. The AWGN pseudoweight is thus a compact expression that measures the "danger" of decoding error due to a specific vertex $x \in \mathscr{P}(H)$.

### 4.4.2 The Fundamental Cone

One simple parameter for estimating the average performance of the LP decoder, for a given code $\mathscr{C}$ and a parity-check matrix $H$, is the *minimal pseudoweight* among the non-zero vertices of $\mathscr{P}(H)$,

$$w_{\mathrm{p,min}}^{\mathrm{AWGN}}(H) = \min\left\{w_{\mathrm{p}}^{\mathrm{AWGN}}(x) : x \neq 0 \text{ is a vertex of } \mathscr{P}(H)\right\},$$

which corresponds to the *most probable* non-zero vertex that accidentally becomes optimal instead of the all-zero one. Note that for an integral vertex $x$ we have $w_{\mathrm{p}}^{\mathrm{AWGN}}(x) = w_{\mathrm{H}}(x)$ (cf. Definition 3.6), which shows that for an ML decoder the minimum Hamming weight takes on this role.

The minimum pseudoweight alone is however still a rather rough estimate of the decoding performance: both the *quantity* of minimum-pseudoweight vertices and the (quantities of the) next larger pseudoweights influence the error probability $P(\text{LP-DECODE}(\lambda) \neq 0)$. Therefore, the *pseudoweight enumerator* of $\mathscr{P}(H)$, i.e., a table containing all occuring pseudoweights of the non-zero vertices alongside with their frequencies, would allow for a better estimation of the decoding performance. Finally, for an *exact* computation of the error rate we would need a description of the region

$$\Lambda = \left\{\lambda \colon \lambda^T x \geq 0 \text{ for all } x \in \mathscr{P}(H)\right\} \tag{4.10}$$

of channel outputs $\lambda$ for which 0 is the optimal solution of (4.4), and then compute the probability $1 - P(\lambda \in \Lambda)$ by integrating the density function given by (4.8) over $\Lambda$. In optimization language, $\Lambda$ is called the *dual cone* of $\mathscr{P}(H)$. See Figure 4.6(a) for an example.

While the three tasks stated above appear to be ascendingly difficult—no efficient algorithm is known to compute the minimum pseudoweight in general—it turns out that $\Lambda$ as defined in (4.10) can be determined by LP duality: assume that the LP decoder (4.4) is given in the form

$$\min \quad \lambda^T x \tag{4.11a}$$
$$\text{s.t.} \quad Ax \leq b, \tag{4.11b}$$

where $A$ and $b$ represent (4.5). The dual of (4.11) is

$$\max \quad -b^T y \tag{4.12a}$$
$$\text{s.t.} \quad A^T y = -\lambda \tag{4.12b}$$
$$y \geq 0. \tag{4.12c}$$

By Theorem 2.11, 0 is optimal for (4.11) if and only if there is an $y$ that is feasible for (4.12) with $b^T y = 0$. Now 0 is feasible for (4.11), hence $b \geq 0$, which together with (4.12c) implies that $y_j = 0$ whenever $b_j \neq 0$ in a solution $y$ with $b^T y = 0$. Taking a closer look at (4.12), we conclude

$$\lambda \in \Lambda \Leftrightarrow (-\lambda) \in \operatorname{conic}\left(\left\{A_{j,\bullet} \colon b_j = 0\right\}\right).$$

Note that $\Lambda$ is also a polyhedron by Theorem 2.5.

As a by-product of the above calculations, it appears that the rows of $Ax \leq b$ for which $b_j \neq 0$ are irrelevant for $\Lambda$ and hence for the question whether the decoder fails or not. It is easy to show that deleting those rows leads to $\operatorname{conic}(\mathscr{P}(H))$, which motivates the following definition (see also Figure 4.6(b)).

**4.8 Definition:** The conic hull of the fundamental polytope,

$$\mathscr{K}(H) = \operatorname{conic}(\mathscr{P}(H)),$$

is called the *fundamental cone* of $H$. By

$$\mathscr{K}_1(H) = \{x \in \mathscr{K}(H) \colon \|x\|_1 = 1\}$$

we denote the intersection of $\mathscr{K}(H)$ with the unit simplex. ◁

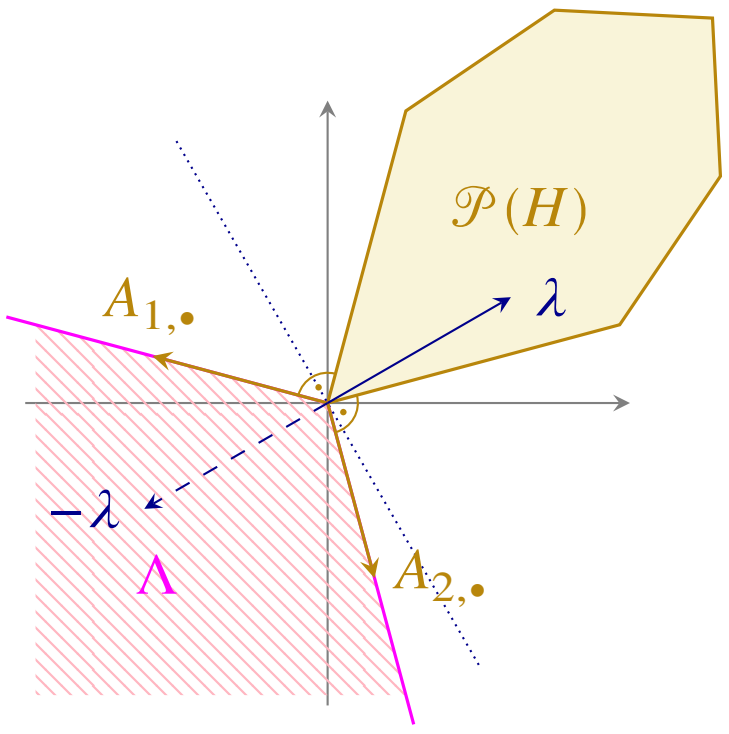

(a) Dual cone $\Lambda$ of $\mathscr{P}(H)$ spanned by two rows $A_{1,\bullet}, A_{2,\bullet}$, and an example $\lambda$ with $-\lambda \in \Lambda$.

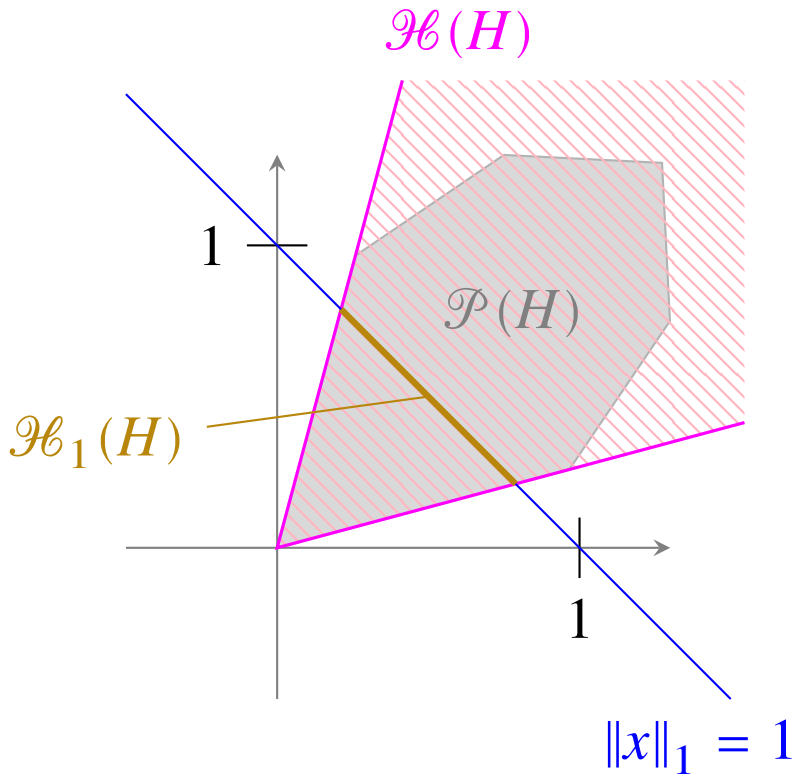

(b) Fundamental cone $\mathscr{K}(H)$ and section $\mathscr{K}_1(H)$ with the unit simplex.

Figure 4.6: Dual cone of an LP (left) and the fundamental cone (right).

From the above discussion we can now formulate the following equivalent conditions for the LP decoder to succeed.

**4.9 Corollary:** *The following are equivalent:*

*(1) The LP decoder correctly decodes $\lambda$ to* $0$.

*(2)* $\lambda \in \Lambda$.

*(3) There is no $x \in \mathcal{K}(H)$ with $\lambda^T x < 0$.*

*(4) There is no $x \in \mathcal{K}_1(H)$ with $\lambda^T x < 0$.* ◁

As a consequence, we can study either of the three sets $\mathcal{P}(H)$, $\mathcal{K}(H)$ or $\mathcal{K}_1(H)$ in order to characterize the LP decoder. Note that while the set $\mathcal{K}(H)$ is larger than $\mathcal{P}(H)$, its description complexity is much lower, because we need only as many forbidden-set inequalities (4.5a) as there are 1-entries in $H$. In addition, observe that the pseudoweight is invariant to scaling, i.e., $w_{\mathrm{p}}^{\mathrm{AWGN}}(\tau x) = w_{\mathrm{p}}^{\mathrm{AWGN}}(x)$ for $\tau > 0$. Consequently, the search for minimum pseudoweight can be restrained to either $\mathcal{K}(H)$ or $\mathcal{K}_1(H)$ as well. For the latter, it takes on the particularly simple form

$$w_{\mathrm{p,min}}^{\mathrm{AWGN}}(H) = \max \left\{ \|x\|_2^2 : x \in \mathcal{K}_1(H) \right\}.$$

While the maximization of $\|\cdot\|_2^2$ over a polytope is NP-hard in general, the most effective algorithms to approach the minimum pseudoweight rely on the above formulation; see [KV03; CS11; RHG14].

### 4.4.3 Graph Covers

There is a fascinating *combinatorial* characterization of the fundamental polytope $\mathcal{P}(H)$ derived from the factor graph $G$ of a parity-check matrix $H$ (cf. Definition 3.8). Central to it is the following definition.

**4.10 Definition (graph cover):** Let $G = (V \cup C, E)$ be the factor graph associated to a parity-check matrix $H$ with variable nodes $V$, check nodes $C$, and edge set $E$. For $m \in \mathbb{N}$, an *m-cover* of $G$ is a factor graph $\bar{G}$ with variable nodes $\bar{V} = V \times \{1, \dots, m\}$, check nodes $\bar{C} = C \times \{1, \dots, m\}$ and a set of $|E|$ permutations $\{e \in \mathbb{S}_M : e \in E\}$ such that the edge set of $\bar{G}$ is

$$\bar{E} = \left\{ (\mathscr{C}_j^{(k)}, x_i^{(l)}) : (\mathscr{C}_j, x_i) = e \in E \text{ and } \pi_e(k) = l \right\},$$

where by $\mathscr{C}_j^{(k)} = (\mathscr{C}_j, k) \in \bar{C}$ we denote the $k$-th copy of $\mathscr{C}_j \in C$ (and $V_i^{(l)}$ analogously). ◁

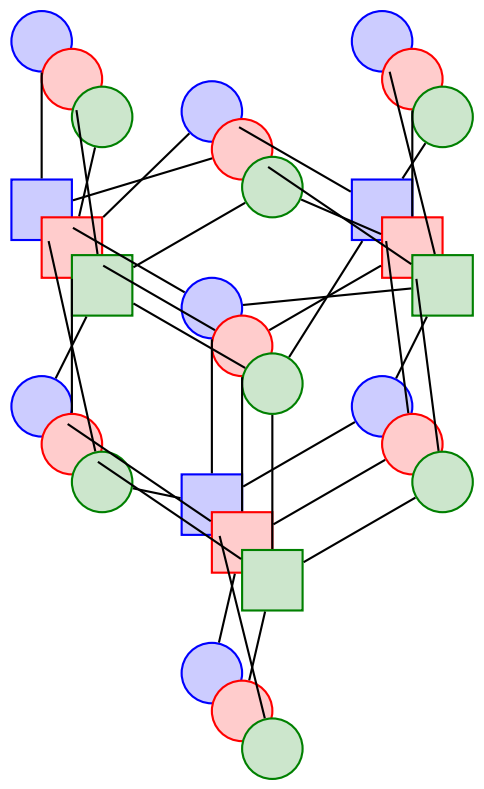

Figure 4.7: A 3-cover of the $(7, 4)$ code shown in Figure 3.5.

Despite the somewhat heavy notation in the above definition, the idea of a graph cover is rather simple: make $m$ identical copies of $G$ and then, for every edge $e = (\mathscr{C}_j, x_i) \in E$, arbitrarily "rewire" the $m$ copies of $e$ in a one-to-one fashion between the copies of $\mathscr{C}_j$ and $x_i$. An example is shown in Figure 4.7.

Since every graph cover of a factor graph $G$ defining a code $\mathscr{C}$ is a factor graph itself, it defines a code $\bar{\mathscr{C}} = \bar{\mathscr{C}}(\bar{G})$ that has $m$ times the block length of $\mathscr{C}$. Let

$$\tilde{x} = (x_1^{(1)}, \dots, x_1^{(m)}, \dots, x_n^{(1)}, \dots, x_n^{(m)}) \in \bar{\mathscr{C}}$$

be such a codeword, where the entries are ordered in the obvious way. Then, the rational $n$-vector $x$ defined by

$$x_i = \frac{1}{m} \sum_{k=1}^{m} x_i^{(k)}$$

is called the *scaled pseudocodeword* of $\mathscr{C}$ associated to $\tilde{x}$. Let $\mathscr{Q}(H)$ denote the union, over all $M > 0$ and all $M$-covers $\bar{G}$ of $G$, of the scaled pseudocodewords associated to all the codewords of $\bar{\mathscr{C}}(\bar{G})$. It then holds that

$$\mathscr{Q}(H) = \mathscr{P}(H) \cap \mathbb{Q}^n,$$

i.e., $\mathscr{Q}(H)$ contains exactly the rational points of $\mathscr{P}(H)$, and hence $\mathscr{P}(H) = \overline{\mathscr{Q}(H)}$.

Graph covers have been proposed in [VK05], among other things, to study the relationship between LP decoding and iterative methods, which can be shown to always compute solutions that are optimal for some graph cover of $G$.

## 4.5 LP Decoding of Turbo Codes

Since one can show that turbo(-like) codes, as introduced in Section 3.5, are special instances of linear block codes, one could compute, for a given turbo code $\mathscr{C}_{\text{TC}}$, a parity-check matrix $H$ defining $\mathscr{C}_{\text{TC}}$, and apply all of the abovementioned theory and algorithms to decode them using linear and integer optimization. In doing so, however, one would neglect the immediate combinatorial structure embodied in $\mathscr{C}$ in virtue of the trellis graphs of the constituent convolutional codes.

In fact, for an individual convolutional code $\mathscr{C}$, it can be shown that ML decoding can be performed by computing a shortest path (due to the simple structure of $T$, this can be achieved in $O(k)$ time) in the trellis $T$ of $\mathscr{C}$, after having introduced a cost value $c_e$ to each trellis edge $e = (v_{i,s}, v_{i+1,s'}) \in E$: as every time step produces $n/k$ output bits, $e$ determines the entries $x^{(i)} = (x_{(i-1)n/k+1}, \dots, x_{in/k}) = x_I$, for an appropriate index set $I$, of the codeword $x$. In view of (4.1), we thus have to define

$$c_e = \sum_{j\colon \ \text{out}(e)_j = 1} \lambda_{I_j}$$

such that the edge cost $c_e$ reflects the portion of the objective function $\lambda^T x$ contributed by including $e$ in the path.

When making the transition to a turbo code $\mathscr{C}_{\text{TC}}$, independently computing a shortest path $P_{\text{a}}$ and $P_{\text{b}}$ in each component trellis $T_{\text{a}}$ and $T_{\text{b}}$, respectively, would not ensure that $P_{\text{a}}$ and $P_{\text{b}}$ are *agreeable*, i.e., fulfill (3.14), and hence match a codeword of $\mathscr{C}_{\text{TC}}$. An ML turbo decoding algorithm would thus need to compute the minimum-cost pair of paths $(P_{\text{a}}, P_{\text{b}})$ in $T_{\text{a}}$ and $T_{\text{b}}$ that additionally is agreeable. While there is no known combinatorial algorithm that efficiently solves such a type of problem (which can be viewed as a generalization of what is called the *equal flow problem* [AKS88]), it is nontheless possible to combine LP decoding with the trellis structure of turbo codes.

One approach is to resort to an LP formulation of the two shortest path problems on $T_{\text{a}}$ and $T_{\text{b}}$ as described in Section 2.4, and then link them by adding linear constraints that represent (3.14): first, write down the constraints of (2.15) for each trellis $T_{\text{a}}$ and $T_{\text{b}}$, where we assume that the decision variable representing an edge $e$ is called $f_e$ instead of $x_e$. Then, for $i = 1, \dots, k$, add an additional contstraint

$$\sum_{\substack{e \in E_i^{\text{a}}:\\ \text{in}(e)=1}} f_e = \sum_{\substack{e \in E_{\pi(i)}^{\text{b}}:\\ \text{in}(e)=1}} f_e$$

to model (3.14), where $E_i^{\text{x}}$ is the edge set of the $i$th segment of trellis $T_{\text{x}}$, $\text{x} \in \{\text{a}, \text{b}\}$. By adding these constraints, the resulting polytope is no longer integral, i.e., the constraints introduce *fractional* vertices that do not correspond to an agreeable pair of paths. Note that there are no $x$ variables for the codeword in the model, since their values are uniquely determined by the values of the $f_e$; it is hence not necessary to include them during the optimization process.

In a similar way as described above (see [HR13b] for the details), a cost value derived from the LLR vector can be assigend to each edge such that the *integer* programming version of the model with additional constraints $f_e \in \{0, 1\}$ is equivalent to ML decoding. Its LP relaxation, called the *turbo LP decoder*, thus has similar properties to the LP decoder from Definition 4.1; in particular, it exhibits the ML certificate property.

While the error-correction performance of the turbo LP decoder has shown to be better than that of the usual LP decoder run on a parity-check matrix representation of the turbo code, solving the turbo LP with a generic LP solver, such as the simplex method, does not exploit the abovementioned possibility to decode the constituent convolutional codes in linear time. Algorithms using this linear-time method as a subroutine to solve the turbo decoding LP efficiently have been proposed in [TRK10; HR13a].